\documentclass[fleqn,10pt,twocolumn]{wlscirep}
\usepackage[utf8]{inputenc}
\usepackage[T1]{fontenc}
\usepackage{balance}
\usepackage[version=4]{mhchem} 
\usepackage{gensymb}
\usepackage{siunitx}
\usepackage{amsmath}
\usepackage{enumerate}
\usepackage{subfigure}
\usepackage{dblfloatfix}

\DeclareSIUnit{\calorie}{cal}
\DeclareSIUnit{\atmosphere}{atm}

\newcommand{\bq}{\begin{equation}} 
\newcommand{\eq}{\end{equation}}
\newcommand\Int{\operatorname{int}}
\newcommand\gas{\operatorname{gas}}
\newcommand\sheet{\operatorname{sheet}}
\newcommand\barrier{\operatorname{barrier}}
\newcommand\TS{\operatorname{TS}}
\newcommand\IIS{\operatorname{SS}}

\newcommand\x{\operatorname{x}}

\newcommand\D{\operatorname{d}}

\title{A new approach to separate hydrogen from carbon dioxide using graphdiyne--like membrane}

\author[1,$\ddagger$,*]{Parham Rezaee}
\author[1,$\ddagger$,$\dagger$]{Hamid Reza Naeij}

\affil[1]{Azadi Ave., P.O.Box 13547-59647, Tehran, Iran}

\affil[*]{Corresponding author: parham.rezaee@alum.sharif.edu}
\affil[$\dagger$]{E-mail: naeij@alum.sharif.edu}
\affil[$\ddagger$]{Recently, graduated from Department of Chemistry at Sharif University of Technology, Tehran, Iran.}

\keywords{Graphdiyne--like membrane, Hydrogen purification, Carbon dioxide capture, DFT calculations, MD simulations, Selectivity and permeance}

\begin{abstract}
In order to separate a mixture of hydrogen (\ce{H2}) and carbon dioxide (\ce{CO2}) gases, we have proposed a new approach employing the graphdiyne--like membrane (GDY--H) using density functional theory (DFT) calculations and molecular dynamics (MD) simulations. GDY--H is constructed by removing one-third diacetylenic (\ce{\bond{-}C\bond{3}C\bond{-}C\bond{3}C\bond{-}}) bonds linkages and replacing with hydrogen atoms in graphdiyne structure. Our DFT calculations exhibit poor selectivity and good permeances for \ce{H2}/\ce{CO2} gases passing through this membrane. To improve the performance of the GDY--H membrane for \ce{H2}/\ce{CO2} separation, we have placed two layers of GDY--H adjacent to each other which the distance between them is 2 \si{nm}. Then, we have inserted 1,3,5-triaminobenzene between two layers. In this approach, the selectivity of \ce{H2}/\ce{CO2} is increased from 5.65 to completely purified \ce{H2} gas. Furthermore, GDY--H membrane represents excellent permeance, about $10^8$ gas permeation unit (GPU), for \ce{H2} molecule at temperatures above 20 K. The \ce{H2} permeance is much higher than the value of the usual industrial limits. Moreover, our proposed approach shows a good balance between the selectivity and permeance parameters for the gas separation which is an essential factor for \ce{H2} purification and \ce{CO2} capture processes in the industry.
\end{abstract}
\begin{document}

\flushbottom
\maketitle
%
%
\thispagestyle{empty}

\section*{Introduction}

Nowadays, \ce{H2} energy is considered as one of the best alternatives to fossil fuels because of its natural abundance, high energy capacity and zero pollutant transpiration \cite{winter_what_2004, andrews_re-envisioning_2012, tollefson_hydrogen_2010, park_hydrogen_2010}. At the \ce{H2} production processes, especially steam--methane reforming reaction, there are many byproducts such as \ce{CO}, \ce{CO2}, \ce{N2} and \ce{CH4} which cause undesirable influences on the energy content and usage of \ce{H2} \cite{alves_overview_2013}. Consequently, developing high--quality and low--cost technologies to separate \ce{H2} from other impurities gases is crucial in the industry \cite{tao_tunable_2014}.

Moreover, \ce{CO2} is regarded as the main greenhouse gas. It is noteworthy that approximately 80\% \ce{CO2} emissions come from the burning of fossil fuels \cite{quadrelli_energyclimate_2007}. It is predicted that the concentration of \ce{CO2} in the atmosphere would increase up to 570 ppm in 2100 which increases the global temperature of about 1.9 \si{\degreeCelsius} \cite{stewart_study_2005}. Therefore, \ce{CO2} capture technology will play an important role in climate change and global warming phenomena \cite{sumida_carbon_2012,samanta_post-combustion_2012,li_carbon_2011}. On the other hand, \ce{CO2} capture is a very expensive technology. So, researchers focus on the development of economical technologies \cite{venna_metal_2015}.

Currently, \ce{H2} separation from \ce{CO2} and \ce{CO2} capture processes have attracted wide attention especially in industrial applications. The common traditional gas separation methods are cryogenic distillation and pressure swing adsorption \cite{sun_application_2015}. However, these methods have disadvantages such as complex performance and high energy consumption. 

So far, many \ce{CO2} capture technologies are used based on physisorption--chemisorption \cite{barzagli_13c_2009,mandal_physical_2005}, amine dry scrubbing \cite{serna-guerrero_new_2008}, metal-organic frameworks (MOFs) \cite{babarao_highly_2010, nandi_single-ligand_2015}, porous organic polymers \cite{zhao_molecular-templating_2019} and ionic liquids \cite{gupta_systematic_2014, budhathoki_molecular_2017}. Recently, membrane-based separation methods are widely used for \ce{H2} purification and \ce{CO2} capture because of low energy consumption, low cost of use and simplicity in performance \cite{jiao_h_2015,david_devlopment_2011,deng_hydrogen_2012,bernardo_membrane_2009}. According to this, various membrane materials such as polymeric membranes \cite{lin_plasticization-enhanced_2006}, MOFs \cite{herm_metalorganic_2011}, nano-porous materials \cite{li_adsorption_2020} and zeolite membranes \cite{li_zeolitic_2010} have been applied to gas separation technology.

The selectivity and permeance are two necessary parameters to investigate the performance of the gas separation membranes. An ideal two-dimensional (2D) membrane would represent a good balance between the selectivity and permeance factors. However, traditional membranes usually have the selectivity-permeance trade-off challenge \cite{robeson_correlation_1991, robeson_upper_2008, gao_versatile_2017}. The permeance is inversely related to the membrane thickness. Hence, one-atom-thick membrane could be an excellent candidate for gas separation \cite{chang_585_2017}.

In the past decade, the design and construction of appropriate 2D membranes for gas separation have dedicated a lot of attention \cite{kang_hydrogen_2009,giraudet_ordered_2010, li_selective_2018}. Recently, carbon allotropes have been used as the gas separation membranes \cite{schrier_fluorinated_2011,liu_selectivity_2015, rezaee_graphenylene1_2020}. These structures show many properties such as high mechanical and chemical stability and periodically distributed uniform pores which make them suitable candidates for the gas separation \cite{jiao_h_2015,bartolomei_graphdiyne_2014}. Among various carbon allotropes, graphdiyne (GDY) is a new 2D carbon allotrope composed of sp and sp$^2$ hybridized carbon atoms which can be constructed by replacing some carbon--carbon bonds in graphene with uniformly distributed diacetylenic linkages \cite{xie_frontispiece:_2020}. This structure was firstly synthesized on the surface of copper using a cross-coupling reaction \cite{li_architecture_2010}. Theoretical and experimental studies show that the existence of sp and sp$^2$ hybridized carbon in GDY leads to high $\pi$--conjunction, wide interplanar spacing, excellent chemical stability, extreme hardness and high thermal resistance of this structure \cite{jia_synthesis_2017, haley_cheminform_2000, sun_graphdiyne:_2015, hui_highly_2019, xue_anchoring_2018, huang_progress_2018, li_graphdiyne_2014}. Furthermore, the heat of formation of GDY is reported about 18.3 kcal per g--atom C, which makes it to be the most stable carbon allotrope containing diacetylenic linkages \cite{haley_carbon_1997}.
  
Many researches have been done to study the gas separation process through the GDY monolayer membrane because of its abundant uniform pores, the size of pores and one-atom thickness. For example, Cranford and Buehler studied the influences of temperature and pressure on \ce{H2} purification from \ce{CO} and \ce{CH4} in the GDY membrane using MD simulations \cite{cranford_selective_2012}. Zhang et al. represented that GDY with larger pores shows a high selectivity for \ce{H2}/ \ce{CH4}, but a relatively low selectivity over small molecules such as \ce{CO} and \ce{N2} \cite{zhang_tunable_2012}. Jiao et al. based on DFT calculations showed that the selectivity of \ce{H2} toward \ce{CH4} and \ce{CO} in the GDY monolayer membrane is much higher than those of silica and carbon membranes \cite{jiao_graphdiyne:_2011}.

It has been proved that changing the pore size of sp--sp$^2$ hybridized carbon in the GDY by substituting some diacetylenic linkages with heteroatoms could be a promising method to improve the performance of the GDY monolayer membrane in the gas separation process \cite{sang_excellent_2017}. In this regard, Desroches et al. synthesized the GDY--like nanoribbons (GDNR) in which one-third diacetylenic linkages of GDNR were substituted with \ce{H} atoms which leads to construct the rhomboidal pores instead of triangular pores \cite{desroches_synthesis_2015}. A nitrogen modified GDY is also investigated concerning its performance for \ce{H2} purification from \ce{CH4} and \ce{CO}. This structure shows high performance for \ce{H2} purification by decreasing \ce{H2} diffusion energy barrier \cite{jiao_h_2015}. Moreover, Zhao et al. designed three GDY--like monolayer membranes by replacing one-third diacetylenic linkages with three heteroatoms \ce{H}, \ce{F} and \ce{O} (GDY--H, GDY--\ce{F} and GDY--\ce{O} membranes, respectively) to control the pore size of GDY for separating a mixture of \ce{CO2}/\ce{N2}/\ce{CH4} gases. Then, they investigated the separation performance of these membranes using DFT calculations and MD simulations. Their study showed that the GDY--H membrane exhibits poor selectivity for \ce{CO2}/\ce{N2}/\ce{CH4} gases, while the GDY--\ce{F} and GDY--\ce{O} membranes can excellently separate \ce{CO2} and \ce{N2} from \ce{CH4} in a wide temperature range \cite{zhao_promising_2017}.

In the present study, we have proposed a new approach to separate \ce{H2} from \ce{CO2} using GDY--H monolayer membrane which designed by Zhao et al. \cite{zhao_promising_2017}. We have calculated the energy barriers of \ce{H2} and \ce{CO2} gases passing through GDY--H monolayer membrane using DFT calculations. Then, we have obtained the selectivity and permeance of the membrane for \ce{H2} and \ce{CO2} gases. Furthermore, we have placed two layers of GDY--H adjacent to each other which the distance between them is 2 \si{nm}. Then, we have inserted 1,3,5-triaminobenzene (1,3,5-TAB) between two layers. The electron pair of \ce{N} atoms in this structure can improve \ce{CO2} capture process. We have performed MD simulations to calculate the selectivity and permeance of the GDY--H membrane for \ce{H2} and \ce{CO2} in three cases: monolayer of membrane, two layers of membrane and two layers of membrane in the presence of 1,3,5-TAB. Our proposed approach shows high selectivity and excellent permeance for separating a mixture of \ce{H2}/\ce{CO2} gases using the GDY--H membrane in the presence of 1,3,5-TAB at different temperatures.

\section*{Computational Methods}

A large 2D sheet \SI[product-units = power]{28.34 x 28.34}{\angstrom} in xy plane including 240 atoms of C and H is formed to exhibit 2D GDY--H monolayer and calculate the energy barrier of the gases diffusing through the membrane and explain the electron density isosurfaces for the molecules interacting with GDY--H monolayer. Isoelectron density surfaces were calculated by the Gaussian 09 program \cite{Frisch} at the B3LYP/6--31G(d) level with D3 correction \cite{tian_expanded_2015}. These surfaces were plotted at isovalues \SI{0.0065}{\elementarycharge\angstrom^{-3}} to describe the interaction between the electron density of the gas and the pore. According to this method, we obtain the potential energy curves of a single \ce{H2} and \ce{CO2} particle when passing through the membrane vertically and horizontally. Based on the barrier energy which was obtained by potential energy curves, we calculate the selectivity and permeance using kinetic theory of gases in the range of 10--600 \si{\kelvin} in our DFT calculations. The equations for calculating permeance and selectivity parameters are explained in detail in section “Results and Discussion”. In addition, the information of \ce{CO2} capture by 1,3,5-TAB was obtained at the B3LYP/6--311++G(d,p) level with D3 correction. 

We have performed MD simulations to analyze \ce{H2} purification using Forcite code in the Material Studio 6.0 software under canonical (NVT) ensemble condition. The range of temperature, 200--600 \si{\kelvin}, was controlled by the Anderson thermostat.  
The information of \ce{H2} purification and \ce{CO2} capture by periodic boundary conditions used in all dimensions. The interatomic interactions between the gases and the carbon-based membranes were described by a condensed-phase optimized molecular potential for atomistic simulation studies (COMPASS) force field \cite{sun_compass:_1998, shan_influence_2012,wu_fluorine-modified_2014,xu_insights_2015}. The cut-off distance of van der Waals interactions was considered as \SI{12.5}{\angstrom}. We have used the Ewald method to investigate the electrostatic interactions. The cubic boxes with the dimensions of $37.55 \times 37.55 \times 37.55$ \si{\angstrom^3} with 200 1,3,5-TAB molecules and 20 \ce{CO2} molecules were considered to study the radial distribution function (RDF) for carbon atoms in \ce{CO2} and nitrogen atoms in 1,3,5-TAB. 

A cubic boxes with the dimensions of $59.0 \times 49.6 \times 100.0$ \si{\angstrom^3} were separated equally along the z-direction with pieces of the GDY--H membrane in order to confirm the QM results. Moreover, cubic boxes with the dimensions of $59.0 \times 49.6 \times 120.0$ \si{\angstrom^3} were trisected along the z-direction with two pieces of GDY--H membranes in the distance of 2 \si{\nano\meter} from each other which are placed at the middle of the box and constructed one gas reservoir in the first part (the gas mixtures involved 200 \ce{H2} and 200 \ce{CO2} molecules), the region contains 75 molecules of 1,3,5-TAB in the middle and the vacuum region on the top side. The carbon atoms on the edge of the GDY--H monolayer were always fixed and all other atoms were fully relaxed (convergence criterion are respectively met: \SI{1d-4}{\kilo\calorie\per\mole} for total energy, \SI{5d-3}{\kilo\calorie\per\mole\per\angstrom} for force and \SI{5d-5}{\angstrom} for displacement). The total time of simulation was \SI{1000}{\pico\second} and Newton’s equations were integrated using \SI{1}{\femto\second} time steps. We have proposed this theoretical method to analyze whether the performance of GDY--H monolayer to purify \ce{H2} in the presence of \ce{CO2} molecules increases or not. According to the diffused gas molecules through monolayer at the end of the simulation time, we have calculated the selectivity, permeance and the probability density distribution in order to evaluate the performance of GDY--H membrane.

\section*{Results and Discussion}

\begin{figure}[t]
\centering
\includegraphics[scale=0.09]{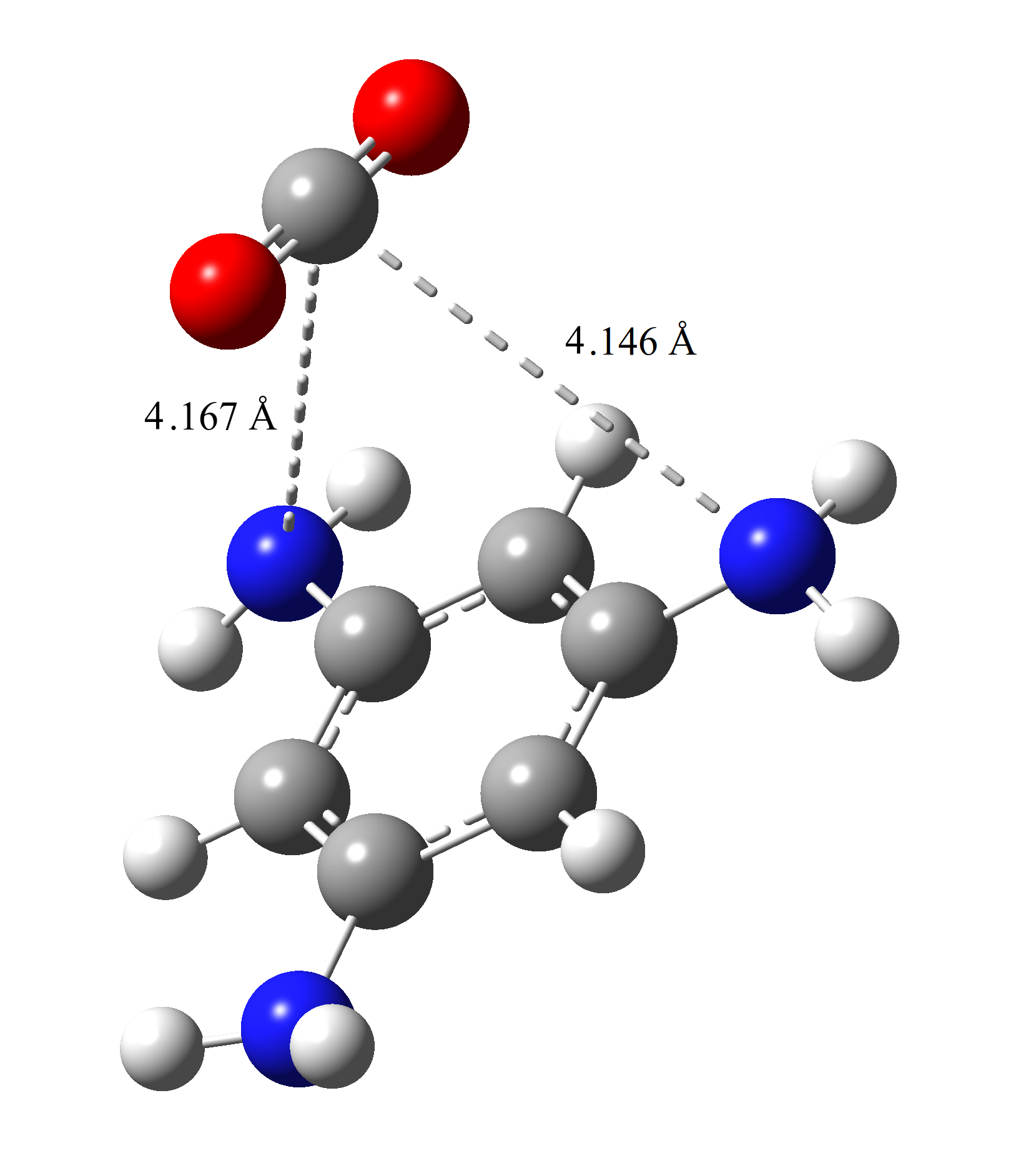}
\caption{Geometry--optimized structure of \ce{CO2} gas in the presence of 1,3,5-TAB molecule.}
\label{capture}
\end{figure}

The stability of the membranes for the gas separation process is an important parameter for their experimental applications. Zhao et al. confirmed the stability of GDY--H monolayer membrane by calculating cohesive energy and phonon dispersion spectra \cite{zhao_promising_2017}. Their results showed that the cohesive energy of GDY is \SI{7.24}{\electronvolt}/atom, which is consistent with theoretical value \SI{7.65}{\electronvolt}/atom \cite{bu_first-principles_2013}. Moreover, the cohesive energy of GDY--H membrane is \SI{6.73}{\electronvolt}/atom \cite{zhao_promising_2017} which is slightly smaller than the value of it for GDY, but is near the $\alpha$--graphyne membrane \SI{6.93}{\electronvolt}/atom \cite{puigdollers_first-principles_2016} and higher than silicene \SI{3.71}{\electronvolt}/atom \cite{li_be_2014}. Therefore, the GDY--H monolayer membrane is strongly bonded structure and rather stable enough for its formation and applications. Moreover, this membrane does not show imaginary frequency in the calculated phonon dispersion spectra \cite{zhao_promising_2017}. It means that the structure of GDY--H membrane is located at the minimum point on the potential energy surfaces. These results indicate that GDY--H membrane could be constructed in the experiments.

Figure \ref{capture} displays the most stable adsorption configurations of \ce{CO2} molecule in the presence of 1,3,5-TAB molecule. For \ce{CO2} molecule, the most stable adsorption sites occurred where C in \ce{CO2} placed at distances of 4.146 and 4.167 \si{\angstrom} toward two nearest N atoms (Figure \ref{capture}) with the binding energy of \SI{0.52}{\electronvolt} and the C--O bond is parallel with C--H bond of the benzene ring in 1,3,5-TAB molecule. Considering entropic penalty, it is expected that the binding energy should be greater than \SI{0.5}{\electronvolt} to effectively capture gas molecules on the solid surfaces. In the float environment of 1,3,5-TAB, we can show that this molecule demonstrates good behavior for the \ce{CO2} capture \cite{guo_co_2015}. 

\begin{table*}[!b]
\large
\centering
  \caption{\ce{H2}/\ce{CO2} selectivities for GDY--H membrane and other proposed membranes at room temperature (\SI{300}{\kelvin}).}
  \label{tbl:QMsel}
  \resizebox{\textwidth}{!}{%
\begin{tabular}{lccccc}
    \hline
Membrane  & GDY--H & $\gamma$--GYN \cite{sang_excellent_2017} & $\gamma$--GYH \cite{sang_excellent_2017} & Graphenylene \cite{song_graphenylene_2013} & g--C$_2$O \cite{zhu_theoretical_2017}\\[10pt]
 & (This work) & & & & \\
    \hline
Selectivity & 5.90 & \num{2d13} & \num{9d17} & \num{1d14} &  \num{3d3}\\
    \hline
  \end{tabular}
  }
\end{table*}

\begin{figure}[t]
\centering
\includegraphics[scale=0.83]{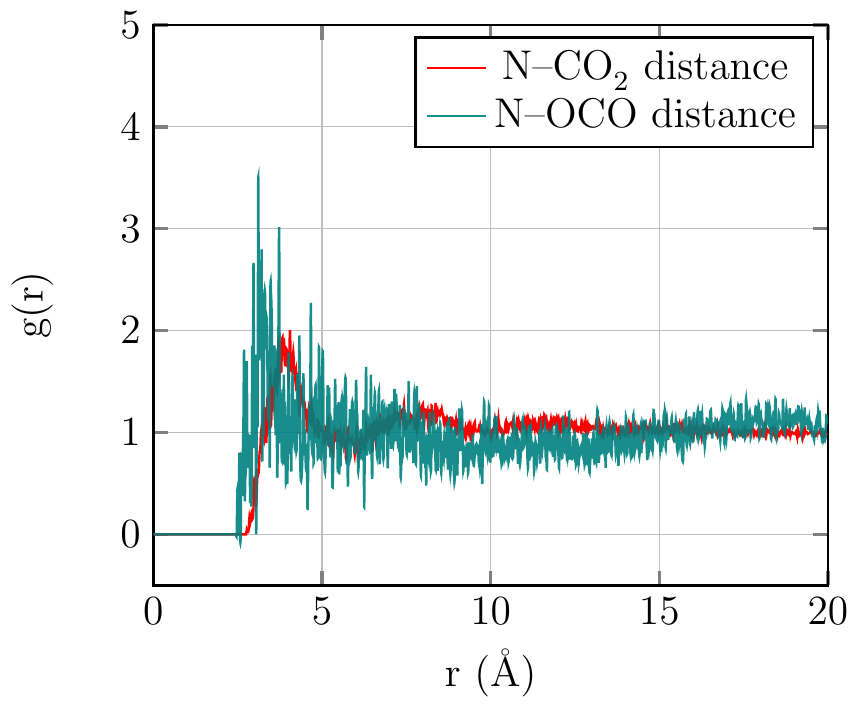}
\caption{Radial distribution functions between N atoms of 1,3,5-TAB and C and O atoms of \ce{CO2}.}
\label{MDcapture}
\end{figure}

Moreover, we performed MD simulation to confirm the results of DFT calculations. We used liquid density to investigate the validity of a proposed force field. In the present study, our results are compared to the experimental data from the other studies. Our results show an appropriate agreement between the predicted density from our force field and the experimental data. The experimental value of the density of the 1,3,5-TAB is 1.279 \si{\centi\metre^3 \per\milli\litre} at 298.15 \si{\kelvin} and 1 bar pressure, and the simulated density is 1.246 \si{\centi\metre^3 \per\milli\litre} which is nearly $\sim$3\% lower than the experimental value. Due to the predictive nature of the calculations, it seems that this level of agreement is suitable. 

The RDF presents information about microstructure considering the nature of interactions as well as the arrangement of the molecules and can be defined as \cite{brehm_travis_2011}
\bq
g_{i,j}(r)={\frac{V}{N_{i}N_{j}}{\sum_{i=1}^{N_i}}{\sum_{j=i+1}^{N_j}}{\langle {\delta (r-| {\overrightarrow{r_{i}}(t) - \overrightarrow{r_{j}}(t)} |)} \rangle}_t}
\eq
where $\overrightarrow{r_{i}}$ and $\overrightarrow{r_{j}}$ denote the position vectors of the $i^{th}$ and the $j^{th}$ particles and the bracket denotes the ensemble average on the distance between atoms $i$ and $j$. Moreover, $N$ and $V$ represent the number of particles and volume, respectively. Each RDF represents the distance-dependent relative probability for observing a given site or atom in relation to some central atom or site. Figure \ref{MDcapture} shows the RDF for the N atom of the 1,3,5-TAB with the C and O atoms of \ce{CO2} molecules. As shown in Figure \ref{MDcapture}, a sharp and intense peak in the RDF is seen for C at about 4.05 \si{\angstrom}. Broader peaks at 7.43 \si{\angstrom} and roughly 11.11 \si{\angstrom} are also seen. Moreover, a sharp and intense peak in the RDF is seen at about 3.73 \si{\angstrom}, indicating the relatively strong interaction between the 1,3,5-TAB and the \ce{CO2} molecules.

\begin{figure}[t]
\centering
\includegraphics[scale=0.85]{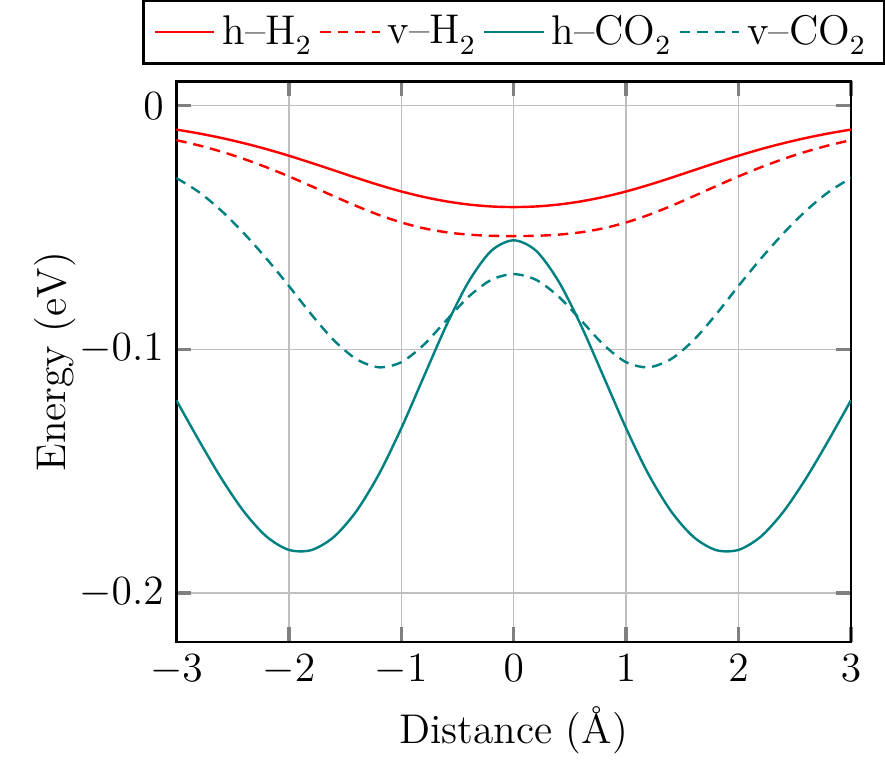}
\caption{Minimum energy pathways for \ce{H2} and \ce{CO2} gases passing horizontally (h) and vertically (v) through GDY--H membrane in the distance \SI{\pm 3}{\angstrom} from the center of the pore.}
\label{BEQM}
\end{figure}

In the gas separation membranes, the interaction energy between the gases and the membrane can be defined as \cite{rezaee_graphenylene1_2020}
\bq
E_{\Int}=E_{\gas + \sheet}-(E_{\gas}+E_{\sheet})
\eq
where $E_{\gas + \sheet}$, $E_{\gas}$ and $E_{\sheet}$ are the total energy of the gas molecule adsorbed on the membrane, the energy of the isolated gas molecule and the energy of the membrane, respectively. In Figure \ref{BEQM}, the minimum energy pathways for \ce{H2} and \ce{CO2} gases passing through GDY--H membrane are plotted in the distance \SI{\pm 3}{\angstrom} from the center of the pore. Since the pore size of the membrane is large, we consider the gases passing horizontally and vertically through the membrane. As shown in Figure \ref{BEQM}, the vertical and horizontal situations have minimum energy pathways for \ce{H2} and \ce{CO2} gases, respectively.

\begin{figure}[t]
\centering
\subfigure[]{
\includegraphics[scale=0.06]{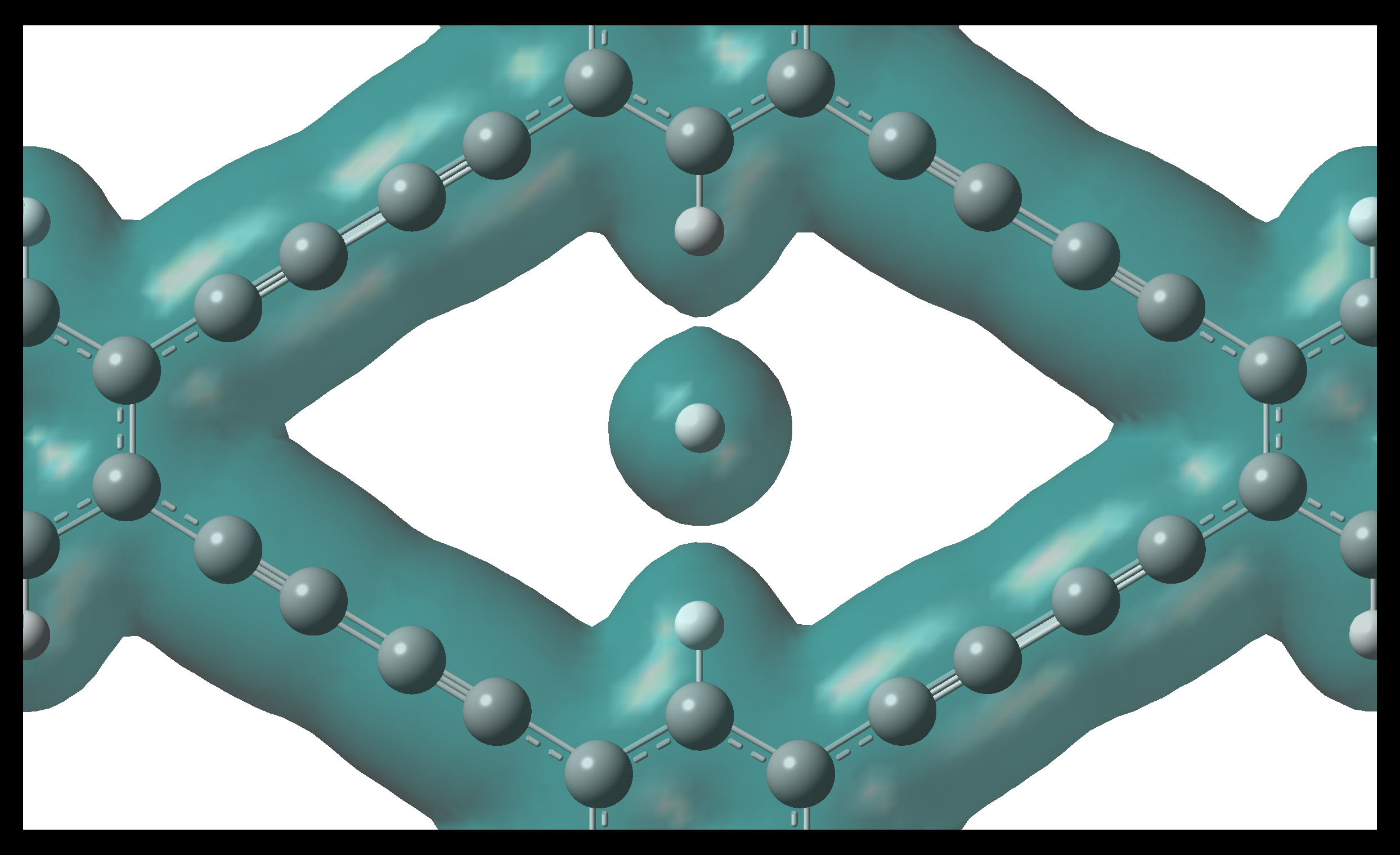}}
\subfigure[]{
\includegraphics[scale=0.063]{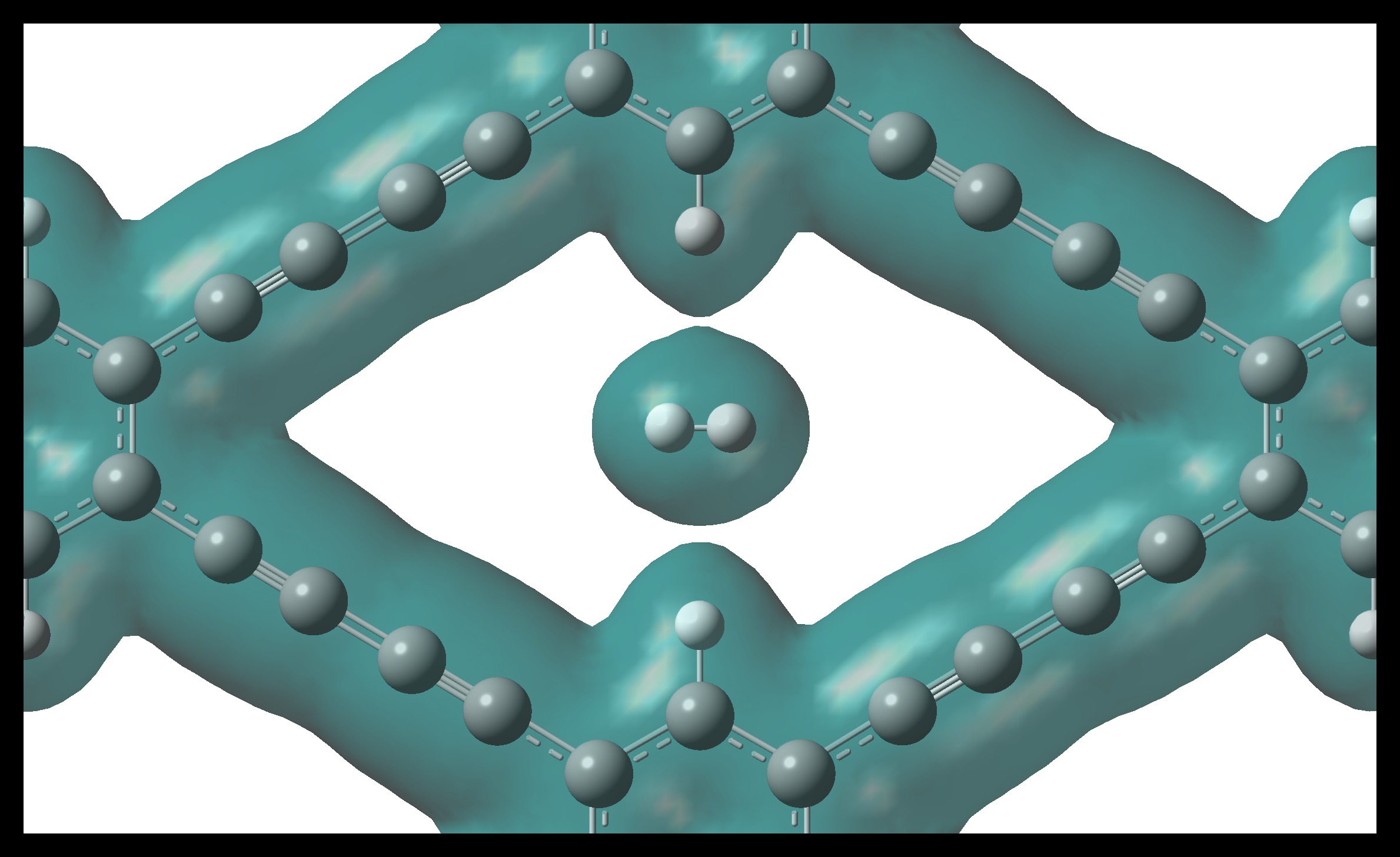}}
\subfigure[]{
\includegraphics[scale=0.062]{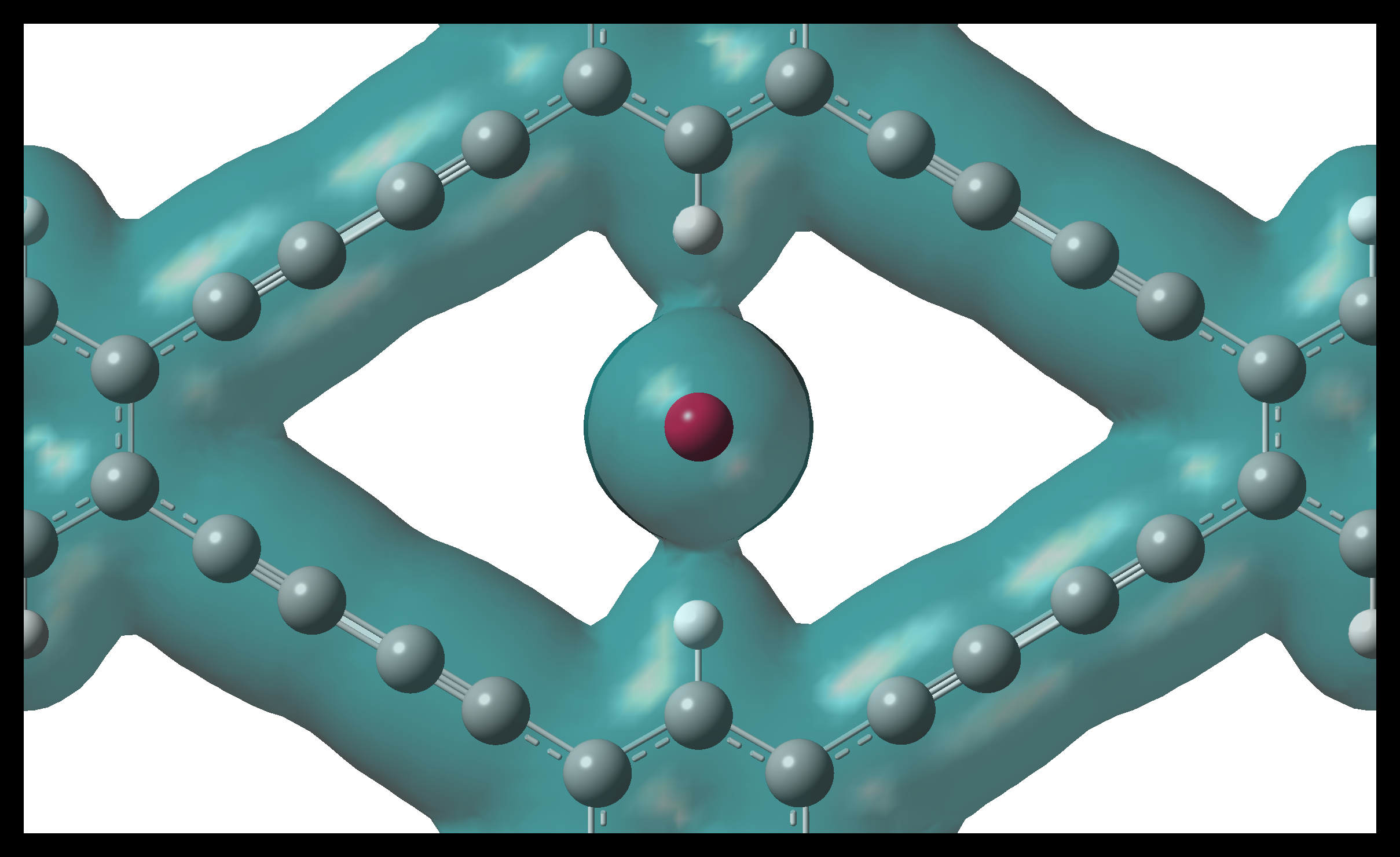}}
\subfigure[]{
\includegraphics[scale=0.06]{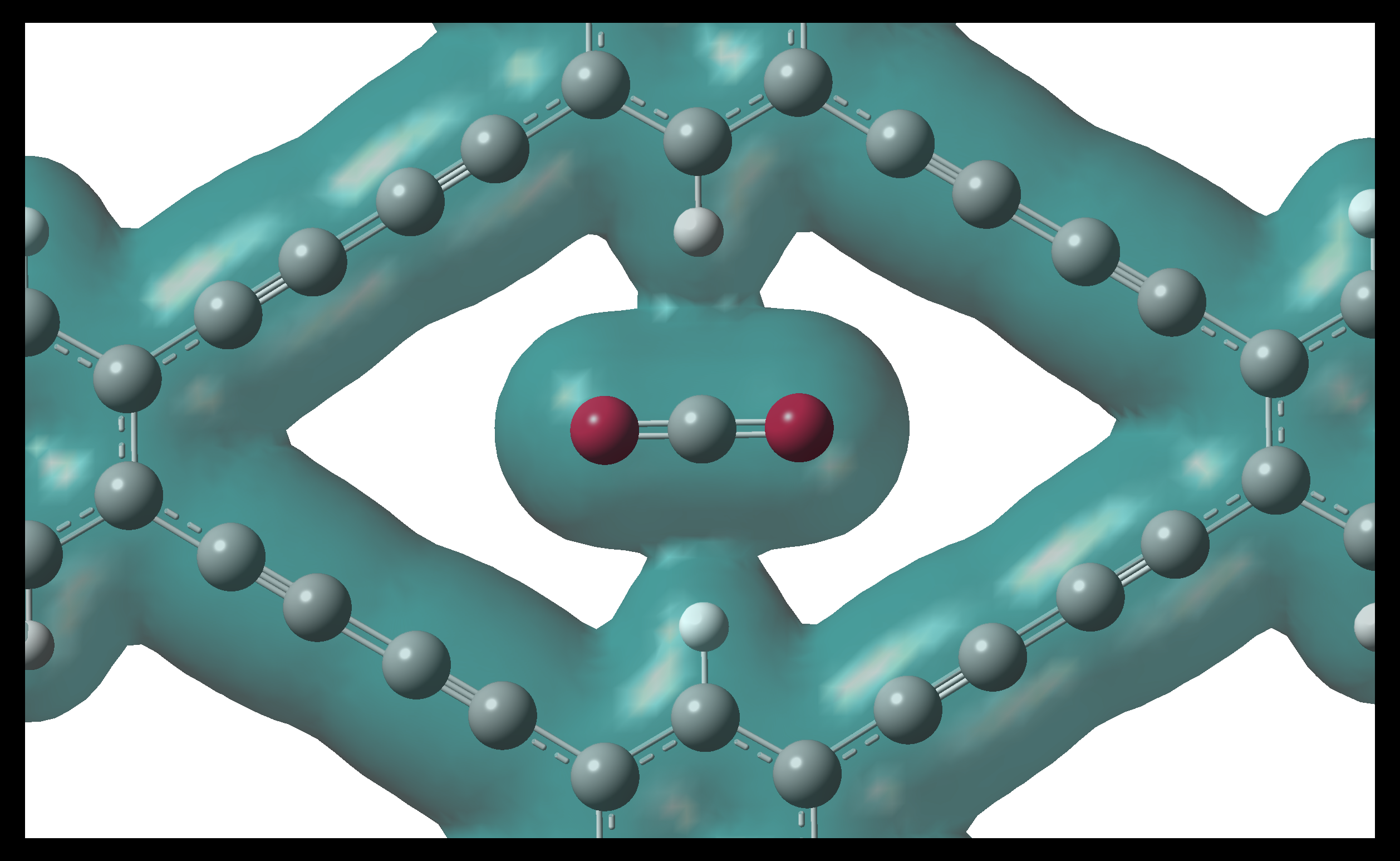}}
\caption{Electron density isosurfaces for \ce{H2} and \ce{CO2} gas molecules passing horizontally (a,c) and vertically (b,d) through GDY--H membrane, respectively. The isovalue is \SI{0.0065} {\elementarycharge\angstrom^{-3}}.}
\label{ISO}
\end{figure}

We also define the diffusion energy barrier for the gases to investigate the process in which the gases passing through the membrane as \cite{rezaee_graphenylene1_2020}
\bq
E_{\barrier}=E_{\TS}-E_{\IIS}
\eq
where $E_{\barrier}$, $E_{\TS}$ and $E_{\IIS}$ represent the diffusion energy barrier, the total energy of the gas molecules and the pore center of the membrane at the transition state and the steady state, respectively. The kinetic diameters ($D_0$) of \ce{H2} and \ce{CO2} gases are 2.60 and 3.30 \si{\angstrom}, respectively and the energy barriers of the gases passing through GDY--H membrane are 0.032 and 0.078 \si{\electronvolt}, respectively.

Furthermore, we have drawn the isoelectron density surfaces at isovalue \SI{0.0065} {\elementarycharge\angstrom^{-3}} in Figure \ref{ISO} to investigate the electron overlaps between \ce{H2} and \ce{CO2} molecules passing horizontally and vertically and GDY--H monolayer membrane. As shown in Figure \ref{ISO}, the energy barrier for \ce{H2} is very low due to the low electron overlap between \ce{H2} and the membrane. On the other hand, more electron overlap between \ce{CO2} molecule and GDY--H membrane makes the higher energy barrier for \ce{CO2} gas.

As we mentioned before, the performance of the gas separation membranes is evaluated by two factors: selectivity and permeance. Here, we investigate these parameters for \ce{H2} and \ce{CO2} gases passing through GDY--H membrane.

We estimate the selectivity of \ce{H2} toward \ce{CO2} passing through GDY--H membrane using the Arrhenius equation which is defined as \cite{sang_excellent_2017}
\bq
S_{{\x} /{\gas}}=\frac{r_{\x}}{r_{\gas}}=\frac{A_{\x} e^{-E_{\x}/RT}}{A_{\gas}e^{-E_{\gas}/RT}}
\eq
where $r$ is the diffusion rate and $A$ is the diffusion prefactor, which is supposed to be the same for all gases ($A$=\SI{1d11}{\per\second}) \cite{sang_excellent_2017}. Furthermore, $E$, $R$ and $T$ are the diffusion energy barrier, the molar gas constant and the temperature of the gases, respectively.

\begin{figure}[t]
\centering
\includegraphics[scale=0.85]{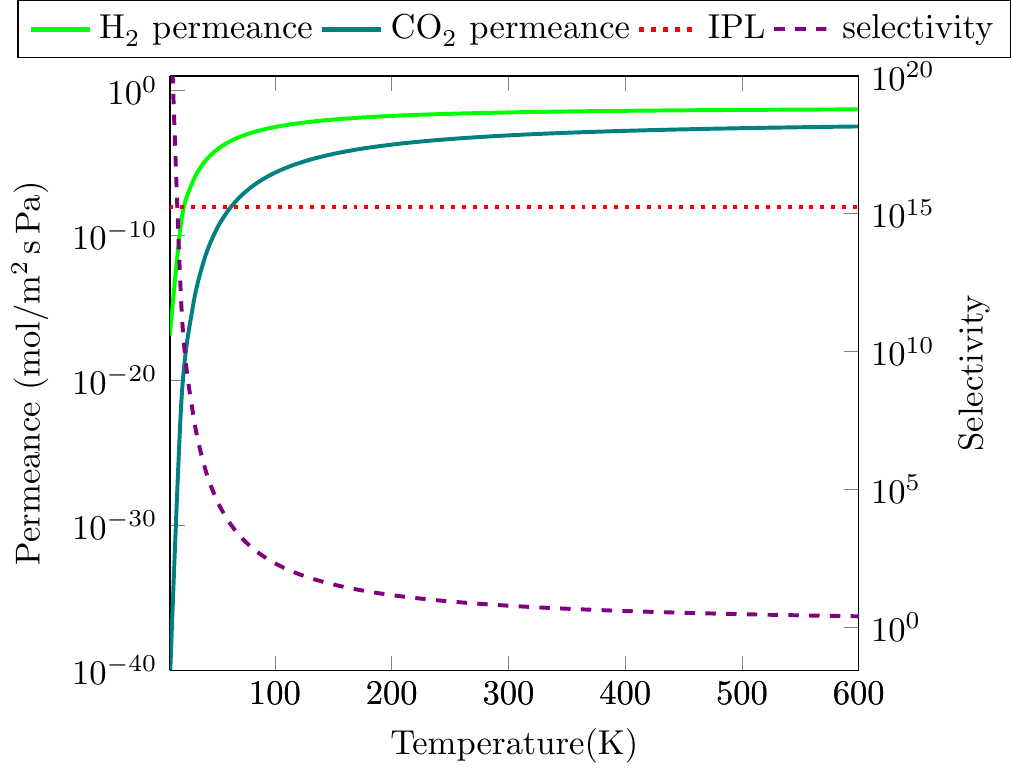}
\caption{Selectivity of \ce{H2}/\ce{CO2} and permeance of \ce{H2} and \ce{CO2} gases passing through GDY--H membrane as a function of temperature based on DFT calculations. The red dotted plot indicates the industrial permeance limit (IPL) for the gas separation process which is 6.7 $\times 10^{-9} \text{mol}/ \si{\metre}^2 \text{s} \text{Pa}$ \cite{sang_excellent_2017}.}
\label{SELPerQM}
\end{figure}

We have drawn the calculated selectivity of GDY--H membrane for \ce{H2} molecule toward \ce{CO2} gas at a wide range of temperatures (\SI{10}{\kelvin}--\SI{600}{\kelvin}) in Figure \ref{SELPerQM}. Our results show that the selectivity for \ce{H2} molecule decreases with increasing temperature. Also, the calculated selectivities of \ce{H2} toward \ce{CO2} for GDY--H membrane and other proposed membranes at room temperature (\SI{300}{\kelvin}) are compared in Table \ref{tbl:QMsel}. As is clear, GDY--H membrane exhibits poor selectivity for \ce{H2} toward \ce{CO2} among the other proposed membranes.

The permeance parameter which indicates the separation efficiency is another important factor to characterize the performance of a gas separation membrane. So, we study the permeance of GDY--H monolayer membrane for separating \ce{H2} from \ce{CO2}.

\begin{figure}
\centering
\subfigure[]{
\includegraphics[scale=0.23]{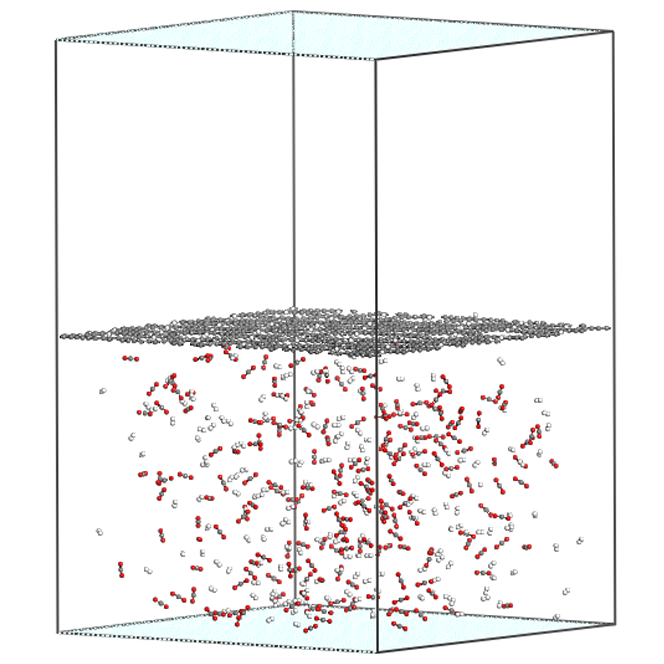}}
\subfigure[]{
\includegraphics[scale=0.27]{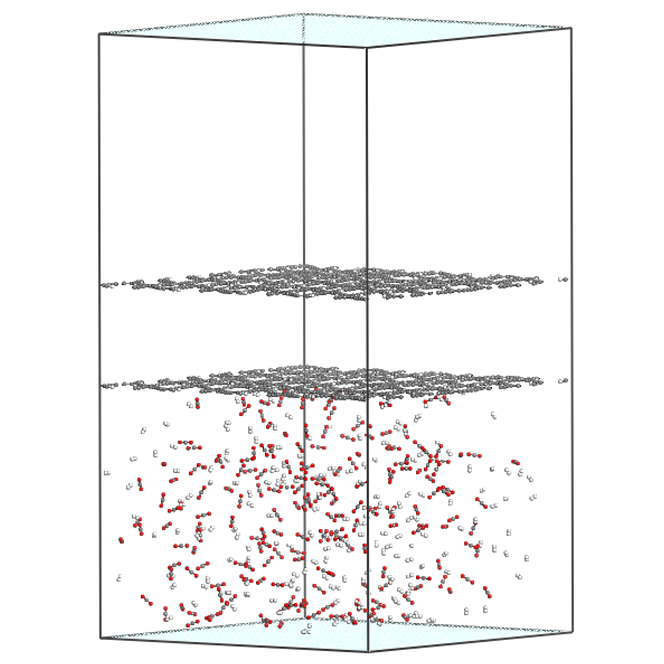}}
\subfigure[]{
\includegraphics[scale=0.27]{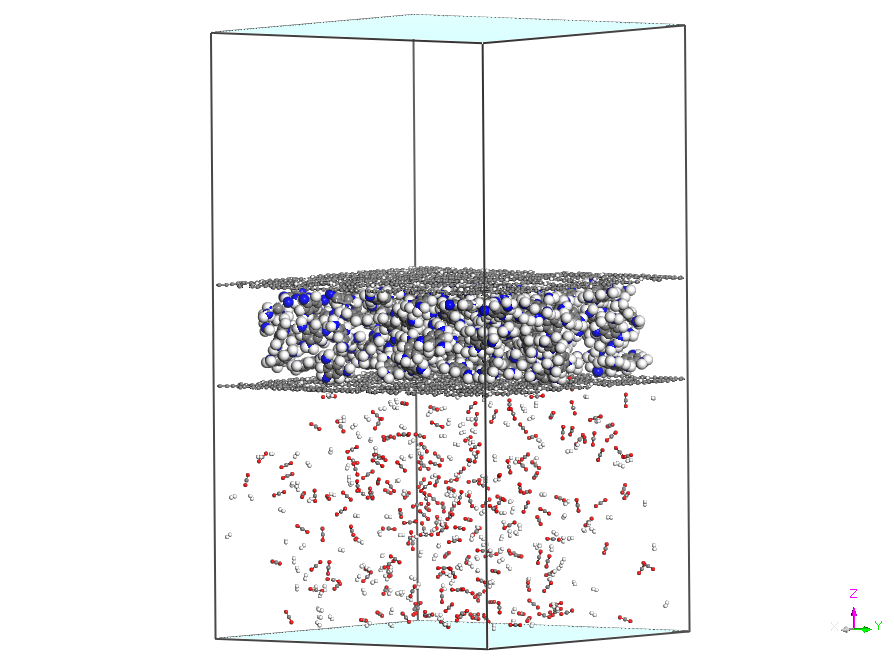}}
\caption{MD simulated configurations of the \ce{H2} and \ce{CO2} gas particles passing through the GDY--H membrane: a) monolayer, b) two layers and c) two layers of the membrane in the presence of 1,3,5-TAB. The height of simulation boxes are 10, 12 and 12 \si{\nano\metre}, respectively.}
\label{MDStr}
\end{figure}

We use the kinetic theory of the gases and the Maxwell--Boltzmann velocity distribution function to analyze the permeances of \ce{H2} and \ce{CO2} gas molecules passing through GDY--H membrane. We define the number of gases colliding with GDY--H sheet as \cite{rezaee_graphenylene1_2020}
\bq
N=\frac{P}{\sqrt{2\pi MRT}}
\eq
where $P$, $M$, $R$ and $T$ are the pressure, here, \num{3d5} Pa, the molar mass, the molar gas constant and the temperature of the gases, respectively. The probability of diffusing of a gas molecule through the pore of the membrane is defined as
\bq
f=\int_{v_B}^{\infty} f(v) \D v
\eq
where $v_B$ and $f(v)$ denote the velocity and the Maxwell velocity distribution function of the gas particles, respectively. The flux of the particles can be expressed as $F = N \times f$. We suppose that the pressure drop $\Delta P$ is \num{1d5} \si{\pascal}. Then, we can express the permeance of the gas molecules passing through the GDY--H membrane as $p=F/\Delta P$ \cite{rezaee_graphenylene1_2020}.

\begin{table}[t]
\large
\centering
  \caption{Number of the gas molecules passing through the GDY--H membrane in the range of 200-600 \si{\kelvin}.}
  \label{tbl:MDnum}
\resizebox{\columnwidth}{!}{
\begin{tabular}{>{\centering\arraybackslash}p{4cm}|>{\centering\arraybackslash}p{1cm}>{\centering\arraybackslash}p{1cm}|>{\centering\arraybackslash}p{1cm}>{\centering\arraybackslash}p{1cm}|>{\centering\arraybackslash}p{1cm}>{\centering\arraybackslash}p{1cm}}
\hline
 & \multicolumn{2}{|c|}{} & \multicolumn{2}{c}{} & \multicolumn{2}{|c}{Two layers}\\
 & \multicolumn{2}{|c|}{Monolayer} & \multicolumn{2}{c}{Two layers} & \multicolumn{2}{|c}{with}\\
 & \multicolumn{2}{|c|}{} & \multicolumn{2}{c}{} & \multicolumn{2}{|c}{1,3,5-TAB}\\
\hline
Temperature (\si{\kelvin}) & \ce{H2} & \ce{CO2} & \ce{H2} & \ce{CO2} & \ce{H2} & \ce{CO2} \\
\hline
200 & 94 & 13 & 79 & 1 & 75 & 0 \\
300 & 96 & 17 & 92 & 2 & 84 & 0 \\
400 & 98 & 23 & 96 & 4 & 88 & 0 \\
500 & 106 & 30 & 98 & 5 & 89 & 0 \\
600 & 106 & 31 & 101 & 7 & 90 & 1 \\
\hline
\end{tabular}
}
\end{table}

\begin{table}[t]
\centering
  \caption{Selectivity of \ce{H2} over \ce{CO2} molecules which passing through the GDY--H membrane at the range of 200--600\si{\kelvin}.}
  \label{tbl:MDsel}
\begin{tabular}{cccc}
\hline
 & & & Two layers \\
Temperature (\si{\kelvin}) & Monolayer & Two layers & with   \\
 & & & 1,3,5-TAB \\
\hline
200 & 7.23 & 79.00 & $\infty$   \\
300 & 5.65 & 46.00 & $\infty$   \\
400 & 4.26 & 24.00 & $\infty$   \\
500 & 3.53 & 19.60 & $\infty$   \\
600 & 3.42 & 14.43 & 90.00   \\
\hline
\end{tabular}
\end{table}

In Figure \ref{SELPerQM}, we have drawn the permeance of the \ce{H2} and \ce{CO2} gases passing through the GDY--H membrane as a function of temperature. The red dotted plot exhibits the industrial permeance limit (IPL) for the gas separation. As shown in Figure \ref{SELPerQM}, with increasing temperature, the permeance of each gas increases largely, while the divergence of permeances between two gases decreases. In other words, by raising the temperature, the kinetic energies ($E=3k_{B}T/2$) of the gases increases. So, the influence of the energy barrier decreases and the gases diffuse through GDY--H membrane more easily. Moreover, it can be concluded that the GDY--H membrane shows the permeance of \ce{H2} and \ce{CO2} gas molecules are much higher than the industrial values at temperatures above \SI{20}{\kelvin} and \SI{80}{\kelvin}, respectively. However, GDY--H membrane does not show an appropriate balance between the selectivity and permeance factors. Therefore, the performance of GDY--H membrane in the separation of \ce{H2} and \ce{CO2} gases is unsuitable.

We now present a new approach to improve the performance of GDY--H membrane for separating a mixture of \ce{H2} and \ce{CO2} gases. 

We place two layers of GDY--H adjacent to each other which the distance between them is 2 \si{\nano\metre}. Then, we insert 1,3,5-TAB between two layers which has a lot of \ce{N} atoms. The electron pair of \ce{N} atoms in this structure can improve \ce{CO2} capture process. We use MD simulations to estimate selectivities and permeances of \ce{H2} and \ce{CO2} gases passing through a monolayer of the membrane, two layers of the membrane and two layers of the membrane in the presence of 1,3,5-TAB at the temperature range of 200--600 \si{\kelvin}.

\begin{table*}[t]
\large
\centering
  \caption{\ce{H2} permeance of the GDY--H membrane in our approach and other proposed membranes at room temperature (\SI{300}{\kelvin}).}
  \label{tbl:MDper}
  \resizebox{\textwidth}{!}{%
\begin{tabular}{lccccccc}
    \hline
 &  &  & Two layers with &  &  & & \\
Membrane & Monolayer & Two layers & 1,3,5-TAB & Graphenylene--1 \cite{rezaee_graphenylene1_2020}& $\gamma$--GYN \cite{sang_excellent_2017} & $\gamma$--GYH \cite{sang_excellent_2017} & g--\ce{C2O} \cite{zhu_theoretical_2017} \\
 & (This work) & (This work) & (This work) & & & &\\
    \hline
Permeance (GPU) & \num{1.06d8} & \num{1.02d8} & \num{9.32d7} & \num{2.6d7} & \num{3.4d7} & \num{1.5d7} & \num{9.4d6}  \\
    \hline
  \end{tabular}
  }
\end{table*}

The MD simulated configurations of the gas particles passing through the porous GDY--H membrane at different temperatures are shown in Figure \ref{MDStr}. The gas molecules adsorb on the surface of the GDY--H monolayer by van der Waals interaction. In the following, they adhere on the surface for a few picoseconds before passing through the monolayer, because of the concentration of the gases is different between the gas reservoir (containing \ce{H2} and \ce{CO2}) and the vacuum space. 

\begin{figure}[t]
\centering
\includegraphics[scale=0.85]{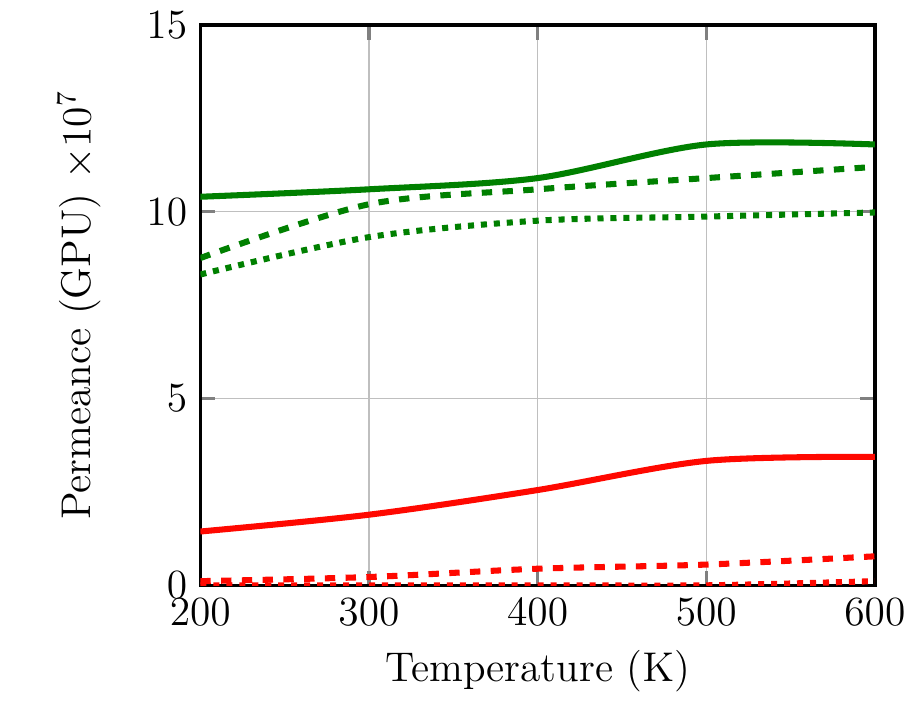}
\caption{Permeance of \ce{H2} and \ce{CO2} gases passing through the GDY--H membrane in our proposed approach as a function of temperature based on MD simulations. The green and red plots represent the permeances of \ce{H2} and \ce{CO2} gases, respectively. The fill plot, dashed plot and dotted plot represent the monolayer, two layers and two layers of the membrane in the presence of 1,3,5-TAB, respectively. 1 GPU=3.35$\times 10^{-10} \text{mol}/ \si{\metre}^2 \text{s} \text{Pa}$ \cite{liu_insights_2013}.}
\label{SELPerMD}
\end{figure}

Based on the MD simulations, one can obtain the numbers of gas molecules passing through the GDY--H membranes after \SI{1}{\nano\second} by counting the number of molecules in the vacuum regions. In this regard, the selectivity of gas A toward gas B can be defined as \cite{wesolowski_pillared_2011}
\begin{align}
S_{A/B} = \frac{x_A/x_B}{y_A/y_B} = \frac{N_A/N_{0,A}}{N_B/N_{0,B}}
\end{align}
where $x_A (x_B)$ and $y_A (y_B)$ are the mole fractions of component A (B) in the vacuum regions and the gas reservoir, respectively and $N_A (N_B)$ and $N_{0,A} (N_{0,B})$ are the corresponding number of molecules A (B).

Furthermore, we can define the permeance of the gases passing through the membrane as \cite{du_separation_2011}
\begin{align}
p=\frac{\nu}{S\times t \times \Delta P}
\end{align}
where $\nu$ and $S$ represent the mole of the gases which diffused through the membrane and the area of the membrane, respectively. Furthermore, $t$ is the time of simulation (\SI{1}{\nano\second}) and the pressure drop ($\Delta P$) is considered 1 \si{\bar} across the pore of GDY--H membrane. 

\begin{figure}
\centering
\subfigure[]{
\includegraphics[scale=0.7]{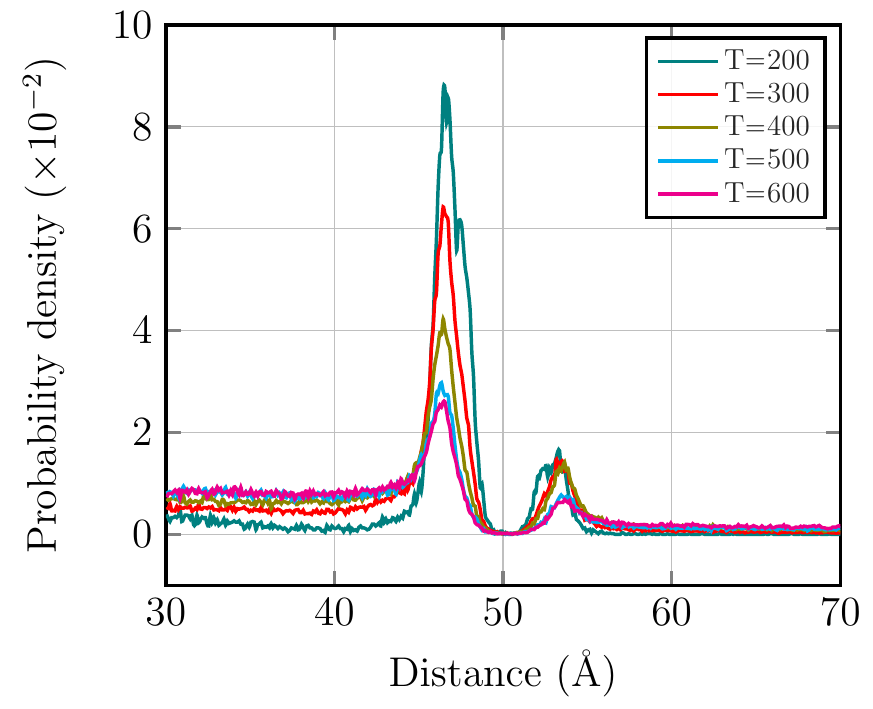}}
\subfigure[]{
\includegraphics[scale=0.7]{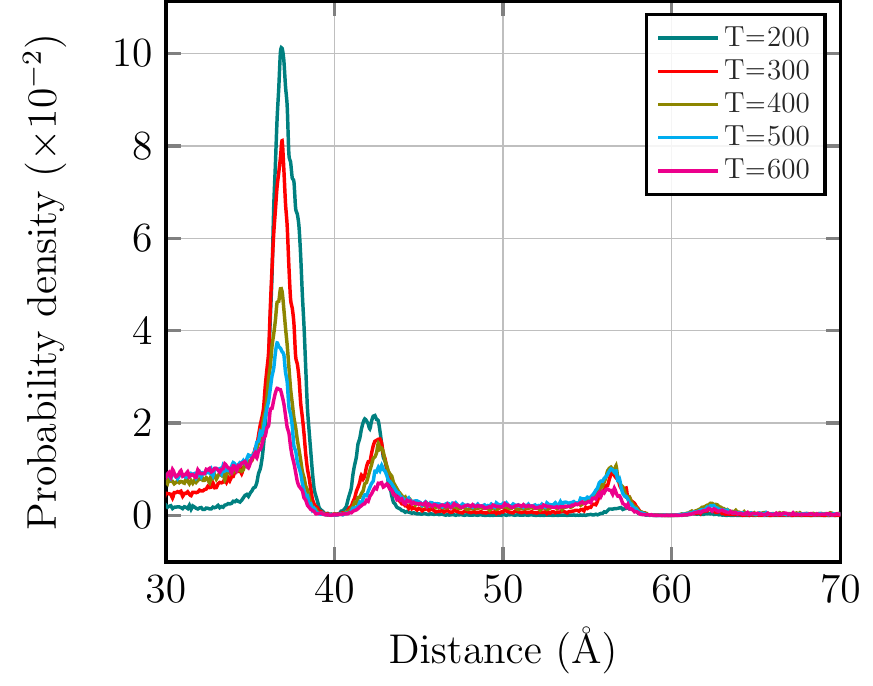}}
\subfigure[]{
\includegraphics[scale=0.7]{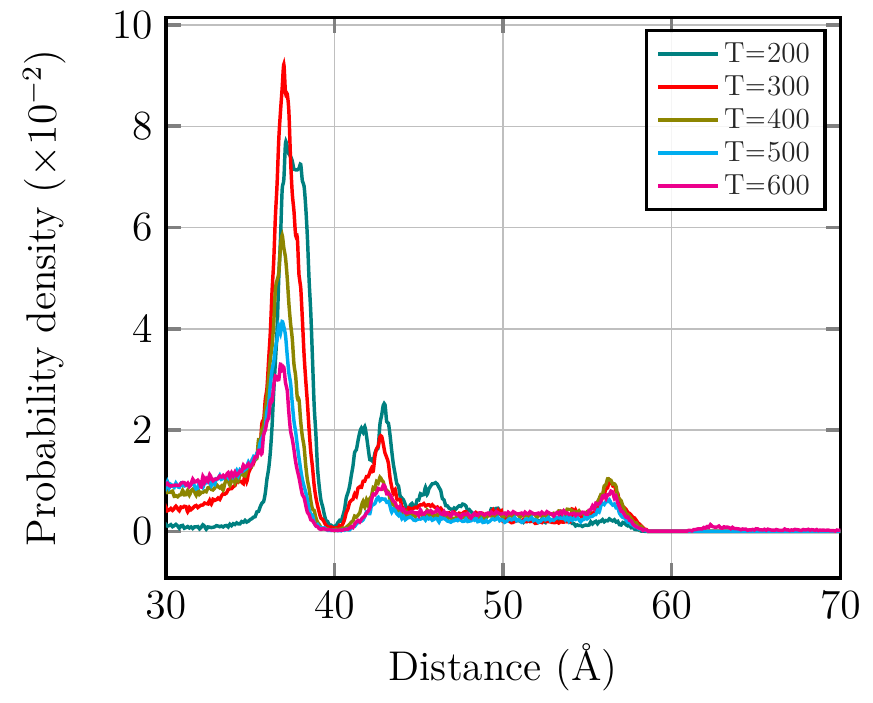}}
\caption{Probability density distribution of the \ce{CO2} molecules passing through the GDY--H membrane at different temperatures as a function of distance in a) monolayer, b) two layers and c) two layers of the membrane in the presence of 1,3,5-TAB.}
\label{Pro}
\end{figure}

The number of gas molecules passing through the monolayer, two layers and two layers of the GDY--H membrane in the presence of 1,3,5-TAB in the range of 200-600 K where given in Table \ref{tbl:MDnum}. In all three cases, as the temperature enhances, the number of particles passing through the membrane increases. However, for two layers and two layers of the GDY--H membrane in the presence of 1,3,5-TAB cases, the passing of \ce{CO2} gases is very negligible. For the third case, it reaches almost zero. This result shows that the presence of 1,3,5-TAB between two layers of GDY--H membrane has been able to capture the \ce{CO2} gas.

The selectivity of \ce{H2} toward \ce{CO2} gases passing through monolayer, two layers and two layers of the GDY--H in the presence of 1,3,5-TAB were given in table \ref{tbl:MDsel}. As it is clear the selectivity of \ce{H2}/\ce{CO2} is increased in the presence of 1,3,5-TAB. Moreover, the permeance of \ce{H2} and \ce{CO2} molecules in the three cases were drawn in Figure \ref{SELPerMD}. It can be seen that the permeance of \ce{H2} and \ce{CO2} gases passing through GDY--H membrane is very high (about $10^8$ GPU). In addition, the permeance of \ce{H2} and \ce{CO2} molecules enhances with increasing temperature. 

The calculated permeances of \ce{H2} for GDY--H membrane in our approach together with that of the previously proposed membrane at room temperature are summarized in Table \ref{tbl:MDper}. As is clear, our approach shows appropriate \ce{H2} permeance for the GDY--H membrane in comparison to the other proposed membrane. The size of the pores in the GDY--H membrane is large in comparison to the other carbon allotrope membranes which leads to a weaker electrostatic and Lennard--Jones interactions between the gas molecules and the membrane. So, the gas separation process will be harder. However, the presence of 1,3,5-TAB in our approach facilitated the \ce{CO2} capture process which leads to improve the selectivity and permeance of the GDY--H membrane. Consequently, our approach shows an appropriate balance between selectivity and permeance factors for the separation of \ce{H2} and \ce{CO2} gases.

Furthermore, the probability density distributions of \ce{CO2} gases as a function of distance to GDY--H membrane
were drawn at different temperatures in Figure \ref{Pro}. As shown in Figure \ref{Pro}, in the monolayer case, we conclude that there is physical adsorption of \ce{CO2} gases on near the membrane. In two layers case, \ce{CO2} gases which passed through the first layer approach to the second layer and adsorb physically in the near of it. In the third case, the probability density of \ce{CO2} gases is increased which shows that \ce{CO2} gases are captured by 1,3,5-TAB. It means that there is no physical adsorption for \ce{CO2} gases (except a single peak at 600 K). These curves exhibit adsorption height for the \ce{CO2} gases in the range of 2-3 \si{\angstrom} from the GDY--H monolayer at low temperatures which is in good agreement with the results obtained by DFT calculations. As the temperature increases, the kinetic energy of the gas particles enhances. Consequently, they overcome the adsorption energy and desorbed from the GDY--H membrane easily. So, the probability distribution for each \ce{CO2} gas decreases at high temperatures.

\section*{Conclusion}

Recent advances in gas separation technology provide new perspectives for the use of carbon allotropes for the development of gas separation membranes. However, one of the main challenges of the most designed carbon membranes is the selectivity-permeance trade-off challenge. Therefore, developing new approaches for the gas separation process based on carbon allotrope membranes seems essential. 

In this work, we proposed a new approach to improve the performance of a GDY--like membrane (GDY--H) to separate a mixture of \ce{H2} and \ce{CO2} gases. This membrane is designed by substituting one--third diacetylenic linkages in GDY structure with hydrogen atoms and the stability of it confirmed by Zhao et al. \cite{zhao_promising_2017}.

First, regarding the calculated energy barriers for the gases, we investigated the performance of GDY--H monolayer membrane for \ce{H2} and \ce{CO2} separation based on DFT calculations. Our results show poor selectivity and good permeance for \ce{H2} and \ce{CO2} gases passing through the membrane. The permeance for \ce{H2} and \ce{CO2} gases are much higher than the value of them in the current industrial applications especially at temperatures above \SI{20}{\kelvin} and \SI{80}{\kelvin}, respectively. However, this monolayer membrane does not show a good balance between the selectivity and permeance factors. 

To improve the performance of GDY--H membrane, we placed two layers of GDY--H adjacent to each other which the distance between them is 2 nm. Then, we inserted 1,3,5-TAB between two layers which the electron pair of \ce{N} atoms in this structure can improve \ce{CO2} capture process. We performed MD simulations to analyze the selectivity and permeance of GDY--H membrane in three cases: a monolayer of the membrane, two layers of the membrane and two layers of the membrane in the presence of 1,3,5-TAB. Our results show that the selectivity of \ce{H2}/\ce{CO2} is increased from 5.65 to purified \ce{H2} gas in the presence of 1,3,5-TAB. Moreover, GDY--H membrane exhibits excellent permeances, more than 10$^8$ GPU, for \ce{H2} and \ce{CO2} gases. Consequently, this proposed approach represented an appropriate balance between the selectivity and permeance factors for \ce{H2} and \ce{CO2} separation.

We hope our proposed approach will be tested by experimental research groups to study \ce{H2} purification and \ce{CO2} capture processes, which are very crucial technologies in the industry.

\section*{Acknowledgements}

The authors would like to thank Dr. Zahra Jamshidi for her comments on the work and useful discussions. They also acknowledge the Computational Spectroscopy Laboratory of Department of Chemistry at Sharif University of Technology for providing computer facilities.

\section*{Author contributions statement}

P. R. proposed the idea and did the calculations and simulations. P. R. and H. R. N. contributed to the development and completion
of the idea, analyzing the results and discussions. P. R. and H. R. N. participated in writing the manuscript.

\section*{Additional information}
The authors declare no competing financial interests.


\begin{thebibliography}{10}

\urlstyle{rm}
\expandafter\ifx\csname url\endcsname\relax
  \def\url#1{\texttt{#1}}\fi
\expandafter\ifx\csname urlprefix\endcsname\relax\def\urlprefix{URL }\fi
\expandafter\ifx\csname doiprefix\endcsname\relax\def\doiprefix{DOI: }\fi
\providecommand{\bibinfo}[2]{#2}
\providecommand{\eprint}[2][]{\url{#2}}

\bibitem{winter_what_2004}
\bibinfo{author}{Winter, M.} \& \bibinfo{author}{Brodd, R.~J.}
\newblock \bibinfo{journal}{\bibinfo{title}{What {are} {batteries}, {fuel}
  {cells}, and {supercapacitors}?}}
\newblock {\emph{\JournalTitle{Chemical Reviews}}}
  \textbf{\bibinfo{volume}{104}}, \bibinfo{pages}{4245--4270},
  \doiprefix\url{10.1021/cr020730k} (\bibinfo{year}{2004}).

\bibitem{andrews_re-envisioning_2012}
\bibinfo{author}{Andrews, J.} \& \bibinfo{author}{Shabani, B.}
\newblock \bibinfo{journal}{\bibinfo{title}{Re-envisioning the role of hydrogen
  in a sustainable energy economy}}.
\newblock {\emph{\JournalTitle{International Journal of Hydrogen Energy}}}
  \textbf{\bibinfo{volume}{37}}, \bibinfo{pages}{1184--1203},
  \doiprefix\url{10.1016/j.ijhydene.2011.09.137} (\bibinfo{year}{2012}).

\bibitem{tollefson_hydrogen_2010}
\bibinfo{author}{Tollefson, J.}
\newblock \bibinfo{journal}{\bibinfo{title}{Hydrogen vehicles: {Fuel} of the
  future?}}
\newblock {\emph{\JournalTitle{Nature}}} \textbf{\bibinfo{volume}{464}},
  \bibinfo{pages}{1262--1264}, \doiprefix\url{10.1038/4641262a}
  (\bibinfo{year}{2010}).

\bibitem{park_hydrogen_2010}
\bibinfo{author}{Park, H.-L.}, \bibinfo{author}{Yi, S.-C.} \&
  \bibinfo{author}{Chung, Y.-C.}
\newblock \bibinfo{journal}{\bibinfo{title}{Hydrogen adsorption on {Li} metal
  in boron-substituted graphene: {An} ab initio approach}}.
\newblock {\emph{\JournalTitle{International Journal of Hydrogen Energy}}}
  \textbf{\bibinfo{volume}{35}}, \bibinfo{pages}{3583--3587},
  \doiprefix\url{10.1016/j.ijhydene.2010.01.073} (\bibinfo{year}{2010}).

\bibitem{alves_overview_2013}
\bibinfo{author}{Alves, H.~J.} \emph{et~al.}
\newblock \bibinfo{journal}{\bibinfo{title}{Overview of hydrogen production
  technologies from biogas and the applications in fuel cells}}.
\newblock {\emph{\JournalTitle{International Journal of Hydrogen Energy}}}
  \textbf{\bibinfo{volume}{38}}, \bibinfo{pages}{5215--5225},
  \doiprefix\url{10.1016/j.ijhydene.2013.02.057} (\bibinfo{year}{2013}).

\bibitem{tao_tunable_2014}
\bibinfo{author}{Tao, Y.} \emph{et~al.}
\newblock \bibinfo{journal}{\bibinfo{title}{Tunable {hydrogen} {separation} in
  {porous} {graphene} {membrane}: {First}-{principle} and {molecular} {dynamic}
  {simulation}}}.
\newblock {\emph{\JournalTitle{ACS Applied Materials \& Interfaces}}}
  \textbf{\bibinfo{volume}{6}}, \bibinfo{pages}{8048--8058},
  \doiprefix\url{10.1021/am4058887} (\bibinfo{year}{2014}).

\bibitem{quadrelli_energyclimate_2007}
\bibinfo{author}{Quadrelli, R.} \& \bibinfo{author}{Peterson, S.}
\newblock \bibinfo{journal}{\bibinfo{title}{The energy--climate challenge:
  {Recent} trends in \ce{CO2} emissions from fuel combustion}}.
\newblock {\emph{\JournalTitle{Energy Policy}}} \textbf{\bibinfo{volume}{35}},
  \bibinfo{pages}{5938--5952}, \doiprefix\url{10.1016/j.enpol.2007.07.001}
  (\bibinfo{year}{2007}).

\bibitem{stewart_study_2005}
\bibinfo{author}{Stewart, C.} \& \bibinfo{author}{Hessami, M.-A.}
\newblock \bibinfo{journal}{\bibinfo{title}{A study of methods of carbon
  dioxide capture and sequestration--the sustainability of a photosynthetic
  bioreactor approach}}.
\newblock {\emph{\JournalTitle{Energy Conversion and Management}}}
  \textbf{\bibinfo{volume}{46}}, \bibinfo{pages}{403--420},
  \doiprefix\url{10.1016/j.enconman.2004.03.009} (\bibinfo{year}{2005}).

\bibitem{sumida_carbon_2012}
\bibinfo{author}{Sumida, K.} \emph{et~al.}
\newblock \bibinfo{journal}{\bibinfo{title}{Carbon {dioxide} {capture} in
  {metal}-{organic} {frameworks}}}.
\newblock {\emph{\JournalTitle{Chemical Reviews}}}
  \textbf{\bibinfo{volume}{112}}, \bibinfo{pages}{724--781},
  \doiprefix\url{10.1021/cr2003272} (\bibinfo{year}{2012}).

\bibitem{samanta_post-combustion_2012}
\bibinfo{author}{Samanta, A.}, \bibinfo{author}{Zhao, A.},
  \bibinfo{author}{Shimizu, G. K.~H.}, \bibinfo{author}{Sarkar, P.} \&
  \bibinfo{author}{Gupta, R.}
\newblock \bibinfo{journal}{\bibinfo{title}{Post-{combustion} \ce{CO2}
  {capture} {using} {solid} {sorbents}: {A} {review}}}.
\newblock {\emph{\JournalTitle{Industrial \& Engineering Chemistry Research}}}
  \textbf{\bibinfo{volume}{51}}, \bibinfo{pages}{1438--1463},
  \doiprefix\url{10.1021/ie200686q} (\bibinfo{year}{2012}).

\bibitem{li_carbon_2011}
\bibinfo{author}{Li, J.-R.} \emph{et~al.}
\newblock \bibinfo{journal}{\bibinfo{title}{Carbon dioxide capture-related gas
  adsorption and separation in metal-organic frameworks}}.
\newblock {\emph{\JournalTitle{Coordination Chemistry Reviews}}}
  \textbf{\bibinfo{volume}{255}}, \bibinfo{pages}{1791--1823},
  \doiprefix\url{10.1016/j.ccr.2011.02.012} (\bibinfo{year}{2011}).

\bibitem{venna_metal_2015}
\bibinfo{author}{Venna, S.~R.} \& \bibinfo{author}{Carreon, M.~A.}
\newblock \bibinfo{journal}{\bibinfo{title}{Metal organic framework membranes
  for carbon dioxide separation}}.
\newblock {\emph{\JournalTitle{Chemical Engineering Science}}}
  \textbf{\bibinfo{volume}{124}}, \bibinfo{pages}{3--19},
  \doiprefix\url{10.1016/j.ces.2014.10.007} (\bibinfo{year}{2015}).

\bibitem{sun_application_2015}
\bibinfo{author}{Sun, C.}, \bibinfo{author}{Wen, B.} \& \bibinfo{author}{Bai,
  B.}
\newblock \bibinfo{journal}{\bibinfo{title}{Application of nanoporous graphene
  membranes in natural gas processing: {Molecular} simulations of \ce{CH4}
  /\ce{CO2} , \ce{CH4} /\ce{H2S} and \ce{CH4} /\ce{N2} separation}}.
\newblock {\emph{\JournalTitle{Chemical Engineering Science}}}
  \textbf{\bibinfo{volume}{138}}, \bibinfo{pages}{616--621},
  \doiprefix\url{10.1016/j.ces.2015.08.049} (\bibinfo{year}{2015}).

\bibitem{barzagli_13c_2009}
\bibinfo{author}{Barzagli, F.}, \bibinfo{author}{Mani, F.} \&
  \bibinfo{author}{Peruzzini, M.}
\newblock \bibinfo{journal}{\bibinfo{title}{A \ce{^13{C}} {NMR} study of the
  carbon dioxide absorption and desorption equilibria by aqueous 2-aminoethanol
  and {N}-methyl-substituted 2-aminoethanol}}.
\newblock {\emph{\JournalTitle{Energy \& Environmental Science}}}
  \textbf{\bibinfo{volume}{2}}, \bibinfo{pages}{322},
  \doiprefix\url{10.1039/b814670e} (\bibinfo{year}{2009}).

\bibitem{mandal_physical_2005}
\bibinfo{author}{Mandal, B.~P.}, \bibinfo{author}{Kundu, M.} \&
  \bibinfo{author}{Bandyopadhyay, S.~S.}
\newblock \bibinfo{journal}{\bibinfo{title}{Physical {solubility} and
  {diffusivity} of \ce{N2O} and \ce{CO2} into {aqueous} {solutions} of
  (2-{amino}-2-methyl-1-propanol + {monoethanolamine}) and ( \textit{{N}}
  -{methyldiethanolamine} + {monoethanolamine})}}.
\newblock {\emph{\JournalTitle{Journal of Chemical \& Engineering Data}}}
  \textbf{\bibinfo{volume}{50}}, \bibinfo{pages}{352--358},
  \doiprefix\url{10.1021/je049826x} (\bibinfo{year}{2005}).

\bibitem{serna-guerrero_new_2008}
\bibinfo{author}{Serna-Guerrero, R.}, \bibinfo{author}{Da’na, E.} \&
  \bibinfo{author}{Sayari, A.}
\newblock \bibinfo{journal}{\bibinfo{title}{New {insights} into the
  {interactions} of \ce{CO2} with {amine}-{functionalized} {silica}}}.
\newblock {\emph{\JournalTitle{Industrial \& Engineering Chemistry Research}}}
  \textbf{\bibinfo{volume}{47}}, \bibinfo{pages}{9406--9412},
  \doiprefix\url{10.1021/ie801186g} (\bibinfo{year}{2008}).

\bibitem{babarao_highly_2010}
\bibinfo{author}{Babarao, R.}, \bibinfo{author}{Eddaoudi, M.} \&
  \bibinfo{author}{Jiang, J.~W.}
\newblock \bibinfo{journal}{\bibinfo{title}{Highly {porous} {ionic}
  \textit{rht} {metal}-{organic} {framework} for \ce{H2} and \ce{CO2} {storage}
  and {separation}: {A} {molecular} {Ssimulation} {study}}}.
\newblock {\emph{\JournalTitle{Langmuir}}} \textbf{\bibinfo{volume}{26}},
  \bibinfo{pages}{11196--11203}, \doiprefix\url{10.1021/la100509g}
  (\bibinfo{year}{2010}).

\bibitem{nandi_single-ligand_2015}
\bibinfo{author}{Nandi, S.} \emph{et~al.}
\newblock \bibinfo{journal}{\bibinfo{title}{A single-ligand ultra-microporous
  {MOF} for precombustion \ce{CO2} capture and hydrogen purification}}.
\newblock {\emph{\JournalTitle{Science Advances}}}
  \textbf{\bibinfo{volume}{1}}, \bibinfo{pages}{e1500421},
  \doiprefix\url{10.1126/sciadv.1500421} (\bibinfo{year}{2015}).

\bibitem{zhao_molecular-templating_2019}
\bibinfo{author}{Zhao, D.} \emph{et~al.}
\newblock \bibinfo{journal}{\bibinfo{title}{A molecular-templating strategy to
  polyamine-incorporated porous organic polymers for unprecedented {CO2}
  capture and separation}}.
\newblock {\emph{\JournalTitle{Science China Materials}}}
  \textbf{\bibinfo{volume}{62}}, \bibinfo{pages}{448--454},
  \doiprefix\url{10.1007/s40843-018-9333-6} (\bibinfo{year}{2019}).

\bibitem{gupta_systematic_2014}
\bibinfo{author}{Gupta, K.~M.} \& \bibinfo{author}{Jiang, J.}
\newblock \bibinfo{journal}{\bibinfo{title}{Systematic {investigation} of
  {nitrile} {based} {ionic} {liquids} for \ce{CO2} {capture}: {A} {combination}
  of {molecular} {simulation} and \textit{ab {initio}} {calculation}}}.
\newblock {\emph{\JournalTitle{The Journal of Physical Chemistry C}}}
  \textbf{\bibinfo{volume}{118}}, \bibinfo{pages}{3110--3118},
  \doiprefix\url{10.1021/jp411434g} (\bibinfo{year}{2014}).

\bibitem{budhathoki_molecular_2017}
\bibinfo{author}{Budhathoki, S.}, \bibinfo{author}{Shah, J.~K.} \&
  \bibinfo{author}{Maginn, E.~J.}
\newblock \bibinfo{journal}{\bibinfo{title}{Molecular {simulation} {study} of
  the {performance} of {supported} {ionic} {liquid} {phase} {materials} for the
  {separation} of {carbon} {dioxide} from {methane} and {hydrogen}}}.
\newblock {\emph{\JournalTitle{Industrial \& Engineering Chemistry Research}}}
  \textbf{\bibinfo{volume}{56}}, \bibinfo{pages}{6775--6784},
  \doiprefix\url{10.1021/acs.iecr.7b00763} (\bibinfo{year}{2017}).

\bibitem{jiao_h_2015}
\bibinfo{author}{Jiao, Y.}, \bibinfo{author}{Du, A.}, \bibinfo{author}{Smith,
  S.~C.}, \bibinfo{author}{Zhu, Z.} \& \bibinfo{author}{Qiao, S.~Z.}
\newblock \bibinfo{journal}{\bibinfo{title}{\ce{H2} purification by
  functionalized graphdiyne - role of nitrogen doping}}.
\newblock {\emph{\JournalTitle{Journal of Materials Chemistry A}}}
  \textbf{\bibinfo{volume}{3}}, \bibinfo{pages}{6767--6771},
  \doiprefix\url{10.1039/C5TA01062D} (\bibinfo{year}{2015}).

\bibitem{david_devlopment_2011}
\bibinfo{author}{David, E.} \& \bibinfo{author}{Kopac, J.}
\newblock \bibinfo{journal}{\bibinfo{title}{Devlopment of palladium/ceramic
  membranes for hydrogen separation}}.
\newblock {\emph{\JournalTitle{International Journal of Hydrogen Energy}}}
  \textbf{\bibinfo{volume}{36}}, \bibinfo{pages}{4498--4506},
  \doiprefix\url{10.1016/j.ijhydene.2010.12.032} (\bibinfo{year}{2011}).

\bibitem{deng_hydrogen_2012}
\bibinfo{author}{Deng, X.}, \bibinfo{author}{Luo, D.}, \bibinfo{author}{Qin,
  C.}, \bibinfo{author}{Qian, X.} \& \bibinfo{author}{Yang, W.}
\newblock \bibinfo{journal}{\bibinfo{title}{Hydrogen isotopes separation using
  frontal displacement chromatography with {Pd}–{Al}2o3 packed column}}.
\newblock {\emph{\JournalTitle{International Journal of Hydrogen Energy}}}
  \textbf{\bibinfo{volume}{37}}, \bibinfo{pages}{10774--10778},
  \doiprefix\url{10.1016/j.ijhydene.2012.04.040} (\bibinfo{year}{2012}).

\bibitem{bernardo_membrane_2009}
\bibinfo{author}{Bernardo, P.}, \bibinfo{author}{Drioli, E.} \&
  \bibinfo{author}{Golemme, G.}
\newblock \bibinfo{journal}{\bibinfo{title}{Membrane {gas} {separation}: {A}
  {review}/{state} of the {art}}}.
\newblock {\emph{\JournalTitle{Industrial \& Engineering Chemistry Research}}}
  \textbf{\bibinfo{volume}{48}}, \bibinfo{pages}{4638--4663},
  \doiprefix\url{10.1021/ie8019032} (\bibinfo{year}{2009}).

\bibitem{lin_plasticization-enhanced_2006}
\bibinfo{author}{Lin, H.}
\newblock \bibinfo{journal}{\bibinfo{title}{Plasticization-{enhanced}
  {hydrogen} {purification} {using} {polymeric} {membranes}}}.
\newblock {\emph{\JournalTitle{Science}}} \textbf{\bibinfo{volume}{311}},
  \bibinfo{pages}{639--642}, \doiprefix\url{10.1126/science.1118079}
  (\bibinfo{year}{2006}).

\bibitem{herm_metalorganic_2011}
\bibinfo{author}{Herm, Z.~R.}, \bibinfo{author}{Swisher, J.~A.},
  \bibinfo{author}{Smit, B.}, \bibinfo{author}{Krishna, R.} \&
  \bibinfo{author}{Long, J.~R.}
\newblock \bibinfo{journal}{\bibinfo{title}{Metal-{organic} {frameworks} as
  {adsorbents} for {hydrogen} {purification} and {precombustion} {carbon}
  {dioxide} {capture}}}.
\newblock {\emph{\JournalTitle{Journal of the American Chemical Society}}}
  \textbf{\bibinfo{volume}{133}}, \bibinfo{pages}{5664--5667},
  \doiprefix\url{10.1021/ja111411q} (\bibinfo{year}{2011}).

\bibitem{li_adsorption_2020}
\bibinfo{author}{Li, L.} \emph{et~al.}
\newblock \bibinfo{journal}{\bibinfo{title}{Adsorption and separation of
  propane/propylene on various {ZIF}-8 polymorphs: {Insights} from {GCMC}
  simulations and the ideal adsorbed solution theory ({IAST})}}.
\newblock {\emph{\JournalTitle{Chemical Engineering Journal}}}
  \textbf{\bibinfo{volume}{386}}, \bibinfo{pages}{123945},
  \doiprefix\url{10.1016/j.cej.2019.123945} (\bibinfo{year}{2020}).

\bibitem{li_zeolitic_2010}
\bibinfo{author}{Li, Y.}, \bibinfo{author}{Liang, F.}, \bibinfo{author}{Bux,
  H.}, \bibinfo{author}{Yang, W.} \& \bibinfo{author}{Caro, J.}
\newblock \bibinfo{journal}{\bibinfo{title}{Zeolitic imidazolate framework
  {ZIF}-7 based molecular sieve membrane for hydrogen separation}}.
\newblock {\emph{\JournalTitle{Journal of Membrane Science}}}
  \textbf{\bibinfo{volume}{354}}, \bibinfo{pages}{48--54},
  \doiprefix\url{10.1016/j.memsci.2010.02.074} (\bibinfo{year}{2010}).

\bibitem{robeson_correlation_1991}
\bibinfo{author}{Robeson, L.~M.}
\newblock \bibinfo{journal}{\bibinfo{title}{Correlation of separation factor
  versus permeability for polymeric membranes}}.
\newblock {\emph{\JournalTitle{Journal of Membrane Science}}}
  \textbf{\bibinfo{volume}{62}}, \bibinfo{pages}{165--185},
  \doiprefix\url{10.1016/0376-7388(91)80060-J} (\bibinfo{year}{1991}).

\bibitem{robeson_upper_2008}
\bibinfo{author}{Robeson, L.~M.}
\newblock \bibinfo{journal}{\bibinfo{title}{The upper bound revisited}}.
\newblock {\emph{\JournalTitle{Journal of Membrane Science}}}
  \textbf{\bibinfo{volume}{320}}, \bibinfo{pages}{390--400},
  \doiprefix\url{10.1016/j.memsci.2008.04.030} (\bibinfo{year}{2008}).

\bibitem{gao_versatile_2017}
\bibinfo{author}{Gao, G.} \emph{et~al.}
\newblock \bibinfo{journal}{\bibinfo{title}{Versatile two-dimensional
  stanene-based membrane for hydrogen purification}}.
\newblock {\emph{\JournalTitle{International Journal of Hydrogen Energy}}}
  \textbf{\bibinfo{volume}{42}}, \bibinfo{pages}{5577--5583},
  \doiprefix\url{10.1016/j.ijhydene.2016.07.119} (\bibinfo{year}{2017}).

\bibitem{chang_585_2017}
\bibinfo{author}{Chang, X.} \emph{et~al.}
\newblock \bibinfo{journal}{\bibinfo{title}{585 divacancy-defective germanene
  as a hydrogen separation membrane: {A} {DFT} study}}.
\newblock {\emph{\JournalTitle{International Journal of Hydrogen Energy}}}
  \textbf{\bibinfo{volume}{42}}, \bibinfo{pages}{24189--24196},
  \doiprefix\url{10.1016/j.ijhydene.2017.08.025} (\bibinfo{year}{2017}).

\bibitem{kang_hydrogen_2009}
\bibinfo{author}{Kang, K.~Y.}, \bibinfo{author}{Lee, B.~I.} \&
  \bibinfo{author}{Lee, J.~S.}
\newblock \bibinfo{journal}{\bibinfo{title}{Hydrogen adsorption on
  nitrogen-doped carbon xerogels}}.
\newblock {\emph{\JournalTitle{Carbon}}} \textbf{\bibinfo{volume}{47}},
  \bibinfo{pages}{1171--1180}, \doiprefix\url{10.1016/j.carbon.2009.01.001}
  (\bibinfo{year}{2009}).

\bibitem{giraudet_ordered_2010}
\bibinfo{author}{Giraudet, S.}, \bibinfo{author}{Zhu, Z.},
  \bibinfo{author}{Yao, X.} \& \bibinfo{author}{Lu, G.}
\newblock \bibinfo{journal}{\bibinfo{title}{Ordered {mesoporous} {carbons}
  {enriched} with {nitrogen}: {Application} to {hydrogen} {storage}}}.
\newblock {\emph{\JournalTitle{The Journal of Physical Chemistry C}}}
  \textbf{\bibinfo{volume}{114}}, \bibinfo{pages}{8639--8645},
  \doiprefix\url{10.1021/jp101119r} (\bibinfo{year}{2010}).

\bibitem{li_selective_2018}
\bibinfo{author}{Li, L.} \emph{et~al.}
\newblock \bibinfo{journal}{\bibinfo{title}{Selective gas diffusion in
  two-dimensional {MXene} lamellar membranes: insights from molecular dynamics
  simulations}}.
\newblock {\emph{\JournalTitle{Journal of Materials Chemistry A}}}
  \textbf{\bibinfo{volume}{6}}, \bibinfo{pages}{11734--11742},
  \doiprefix\url{10.1039/C8TA03701A} (\bibinfo{year}{2018}).

\bibitem{schrier_fluorinated_2011}
\bibinfo{author}{Schrier, J.}
\newblock \bibinfo{journal}{\bibinfo{title}{Fluorinated and {nanoporous}
  {graphene} {materials} {as} {sorbents} for {gas} {separations}}}.
\newblock {\emph{\JournalTitle{ACS Applied Materials \& Interfaces}}}
  \textbf{\bibinfo{volume}{3}}, \bibinfo{pages}{4451--4458},
  \doiprefix\url{10.1021/am2011349} (\bibinfo{year}{2011}).

\bibitem{liu_selectivity_2015}
\bibinfo{author}{Liu, H.}, \bibinfo{author}{Chen, Z.}, \bibinfo{author}{Dai,
  S.} \& \bibinfo{author}{Jiang, D.-e.}
\newblock \bibinfo{journal}{\bibinfo{title}{Selectivity trend of gas separation
  through nanoporous graphene}}.
\newblock {\emph{\JournalTitle{Journal of Solid State Chemistry}}}
  \textbf{\bibinfo{volume}{224}}, \bibinfo{pages}{2--6},
  \doiprefix\url{10.1016/j.jssc.2014.01.030} (\bibinfo{year}{2015}).

\bibitem{rezaee_graphenylene1_2020}
\bibinfo{author}{Rezaee, P.} \& \bibinfo{author}{Naeij, H.~R.}
\newblock \bibinfo{journal}{\bibinfo{title}{Graphenylene--1 membrane: {An}
  excellent candidate for hydrogen purification and helium separation}}.
\newblock {\emph{\JournalTitle{Carbon}}} \textbf{\bibinfo{volume}{157}},
  \bibinfo{pages}{779--787}, \doiprefix\url{10.1016/j.carbon.2019.10.064}
  (\bibinfo{year}{2020}).

\bibitem{bartolomei_graphdiyne_2014}
\bibinfo{author}{Bartolomei, M.} \emph{et~al.}
\newblock \bibinfo{journal}{\bibinfo{title}{Graphdiyne {pores}: “{Ad}
  {Hoc}” {openings} for {helium} {separation} {applications}}}.
\newblock {\emph{\JournalTitle{The Journal of Physical Chemistry C}}}
  \textbf{\bibinfo{volume}{118}}, \bibinfo{pages}{29966--29972},
  \doiprefix\url{10.1021/jp510124e} (\bibinfo{year}{2014}).

\bibitem{xie_frontispiece:_2020}
\bibinfo{author}{Xie, C.}, \bibinfo{author}{Wang, N.}, \bibinfo{author}{Li,
  X.}, \bibinfo{author}{Xu, G.} \& \bibinfo{author}{Huang, C.}
\newblock \bibinfo{journal}{\bibinfo{title}{Frontispiece: {Research} on the
  {preparation} of {graphdiyne} and {its} {derivatives}}}.
\newblock {\emph{\JournalTitle{Chemistry -- A European Journal}}}
  \textbf{\bibinfo{volume}{26}}, \bibinfo{pages}{chem.202080361},
  \doiprefix\url{10.1002/chem.202080361} (\bibinfo{year}{2020}).

\bibitem{li_architecture_2010}
\bibinfo{author}{Li, G.} \emph{et~al.}
\newblock \bibinfo{journal}{\bibinfo{title}{Architecture of graphdiyne
  nanoscale films}}.
\newblock {\emph{\JournalTitle{Chemical Communications}}}
  \textbf{\bibinfo{volume}{46}}, \bibinfo{pages}{3256},
  \doiprefix\url{10.1039/b922733d} (\bibinfo{year}{2010}).

\bibitem{jia_synthesis_2017}
\bibinfo{author}{Jia, Z.} \emph{et~al.}
\newblock \bibinfo{journal}{\bibinfo{title}{Synthesis and {properties} of 2d
  {carbon}--{graphdiyne}}}.
\newblock {\emph{\JournalTitle{Accounts of Chemical Research}}}
  \textbf{\bibinfo{volume}{50}}, \bibinfo{pages}{2470--2478},
  \doiprefix\url{10.1021/acs.accounts.7b00205} (\bibinfo{year}{2017}).

\bibitem{haley_cheminform_2000}
\bibinfo{author}{Haley, M.~M.} \& \bibinfo{author}{Wan, W.~B.}
\newblock \bibinfo{journal}{\bibinfo{title}{{ChemInform} {abstract}: {Natural}
  and {non}-{natural} {planar} {carbon} {networks}: {From} {monomeric} {models}
  to {oligomeric} {substructures}}}.
\newblock {\emph{\JournalTitle{ChemInform}}} \textbf{\bibinfo{volume}{31}},
  \bibinfo{pages}{no--no}, \doiprefix\url{10.1002/chin.200038292}
  (\bibinfo{year}{2000}).

\bibitem{sun_graphdiyne:_2015}
\bibinfo{author}{Sun, L.} \emph{et~al.}
\newblock \bibinfo{journal}{\bibinfo{title}{Graphdiyne: {A} two-dimensional
  thermoelectric material with high figure of merit}}.
\newblock {\emph{\JournalTitle{Carbon}}} \textbf{\bibinfo{volume}{90}},
  \bibinfo{pages}{255--259}, \doiprefix\url{10.1016/j.carbon.2015.04.037}
  (\bibinfo{year}{2015}).

\bibitem{hui_highly_2019}
\bibinfo{author}{Hui, L.} \emph{et~al.}
\newblock \bibinfo{journal}{\bibinfo{title}{Highly {Efficient} and {Selective}
  {Generation} of {Ammonia} and {Hydrogen} on a {Graphdiyne}-{Based}
  {Catalyst}}}.
\newblock {\emph{\JournalTitle{Journal of the American Chemical Society}}}
  \textbf{\bibinfo{volume}{141}}, \bibinfo{pages}{10677--10683},
  \doiprefix\url{10.1021/jacs.9b03004} (\bibinfo{year}{2019}).

\bibitem{xue_anchoring_2018}
\bibinfo{author}{Xue, Y.} \emph{et~al.}
\newblock \bibinfo{journal}{\bibinfo{title}{Anchoring zero valence single atoms
  of nickel and iron on graphdiyne for hydrogen evolution}}.
\newblock {\emph{\JournalTitle{Nature Communications}}}
  \textbf{\bibinfo{volume}{9}}, \bibinfo{pages}{1460},
  \doiprefix\url{10.1038/s41467-018-03896-4} (\bibinfo{year}{2018}).

\bibitem{huang_progress_2018}
\bibinfo{author}{Huang, C.} \emph{et~al.}
\newblock \bibinfo{journal}{\bibinfo{title}{Progress in {Research} into {2D}
  {Graphdiyne}-{Based} {Materials}}}.
\newblock {\emph{\JournalTitle{Chemical Reviews}}}
  \textbf{\bibinfo{volume}{118}}, \bibinfo{pages}{7744--7803},
  \doiprefix\url{10.1021/acs.chemrev.8b00288} (\bibinfo{year}{2018}).

\bibitem{li_graphdiyne_2014}
\bibinfo{author}{Li, Y.}, \bibinfo{author}{Xu, L.}, \bibinfo{author}{Liu, H.}
  \& \bibinfo{author}{Li, Y.}
\newblock \bibinfo{journal}{\bibinfo{title}{Graphdiyne and graphyne: from
  theoretical predictions to practical construction}}.
\newblock {\emph{\JournalTitle{Chemical Society Reviews}}}
  \textbf{\bibinfo{volume}{43}}, \bibinfo{pages}{2572},
  \doiprefix\url{10.1039/c3cs60388a} (\bibinfo{year}{2014}).

\bibitem{haley_carbon_1997}
\bibinfo{author}{Haley, M.~M.}, \bibinfo{author}{Brand, S.~C.} \&
  \bibinfo{author}{Pak, J.~J.}
\newblock \bibinfo{journal}{\bibinfo{title}{Carbon {networks} {based} on
  {dehydrobenzoannulenes}: {Synthesis} of {graphdiyne} {substructures}}}.
\newblock {\emph{\JournalTitle{Angewandte Chemie International Edition in
  English}}} \textbf{\bibinfo{volume}{36}}, \bibinfo{pages}{836--838},
  \doiprefix\url{10.1002/anie.199708361} (\bibinfo{year}{1997}).

\bibitem{cranford_selective_2012}
\bibinfo{author}{Cranford, S.~W.} \& \bibinfo{author}{Buehler, M.~J.}
\newblock \bibinfo{journal}{\bibinfo{title}{Selective hydrogen purification
  through graphdiyne under ambient temperature and pressure}}.
\newblock {\emph{\JournalTitle{Nanoscale}}} \textbf{\bibinfo{volume}{4}},
  \bibinfo{pages}{4587}, \doiprefix\url{10.1039/c2nr30921a}
  (\bibinfo{year}{2012}).

\bibitem{zhang_tunable_2012}
\bibinfo{author}{Zhang, H.} \emph{et~al.}
\newblock \bibinfo{journal}{\bibinfo{title}{Tunable {hydrogen} {separation} in
  $sp-sp^2$ {carbon} {membranes}: {A} {first}-{principles} {prediction}}}.
\newblock {\emph{\JournalTitle{The Journal of Physical Chemistry C}}}
  \textbf{\bibinfo{volume}{116}}, \bibinfo{pages}{16634--16638},
  \doiprefix\url{10.1021/jp304908p} (\bibinfo{year}{2012}).

\bibitem{jiao_graphdiyne:_2011}
\bibinfo{author}{Jiao, Y.} \emph{et~al.}
\newblock \bibinfo{journal}{\bibinfo{title}{Graphdiyne: a versatile
  nanomaterial for electronics and hydrogen purification}}.
\newblock {\emph{\JournalTitle{Chemical Communications}}}
  \textbf{\bibinfo{volume}{47}}, \bibinfo{pages}{11843},
  \doiprefix\url{10.1039/c1cc15129k} (\bibinfo{year}{2011}).

\bibitem{sang_excellent_2017}
\bibinfo{author}{Sang, P.} \emph{et~al.}
\newblock \bibinfo{journal}{\bibinfo{title}{Excellent membranes for hydrogen
  purification: {Dumbbell}-shaped porous $\gamma$-graphynes}}.
\newblock {\emph{\JournalTitle{International Journal of Hydrogen Energy}}}
  \textbf{\bibinfo{volume}{42}}, \bibinfo{pages}{5168--5176},
  \doiprefix\url{10.1016/j.ijhydene.2016.11.158} (\bibinfo{year}{2017}).

\bibitem{desroches_synthesis_2015}
\bibinfo{author}{Desroches, M.}, \bibinfo{author}{Courtemanche, M.-A.},
  \bibinfo{author}{Rioux, G.} \& \bibinfo{author}{Morin, J.-F.}
\newblock \bibinfo{journal}{\bibinfo{title}{Synthesis and {properties} of
  {rhomboidal} {macrocyclic} {subunits} of {graphdiyne}-{like} {nanoribbons}}}.
\newblock {\emph{\JournalTitle{The Journal of Organic Chemistry}}}
  \textbf{\bibinfo{volume}{80}}, \bibinfo{pages}{10634--10642},
  \doiprefix\url{10.1021/acs.joc.5b01752} (\bibinfo{year}{2015}).

\bibitem{zhao_promising_2017}
\bibinfo{author}{Zhao, L.} \emph{et~al.}
\newblock \bibinfo{journal}{\bibinfo{title}{Promising monolayer membranes for
  \ce{CO2}/\ce{N2}/\ce{CH4} separation: {Graphdiynes} modified respectively
  with hydrogen, fluorine, and oxygen atoms}}.
\newblock {\emph{\JournalTitle{Applied Surface Science}}}
  \textbf{\bibinfo{volume}{405}}, \bibinfo{pages}{455--464},
  \doiprefix\url{10.1016/j.apsusc.2017.02.054} (\bibinfo{year}{2017}).

\bibitem{Frisch}
\bibinfo{author}{Frisch, A.}
\newblock \bibinfo{journal}{\bibinfo{title}{Gaussian 09w reference}}.
\newblock {\emph{\JournalTitle{Wallingford}}} \bibinfo{pages}{25}
  (\bibinfo{year}{2009}).

\bibitem{tian_expanded_2015}
\bibinfo{author}{Tian, Z.}, \bibinfo{author}{Dai, S.} \&
  \bibinfo{author}{Jiang, D.-e.}
\newblock \bibinfo{journal}{\bibinfo{title}{Expanded {porphyrins} as
  {two}-{dimensional} {porous} {membranes} for \ce{CO2} {separation}}}.
\newblock {\emph{\JournalTitle{ACS Applied Materials \& Interfaces}}}
  \textbf{\bibinfo{volume}{7}}, \bibinfo{pages}{13073--13079},
  \doiprefix\url{10.1021/acsami.5b03275} (\bibinfo{year}{2015}).

\bibitem{sun_compass:_1998}
\bibinfo{author}{Sun, H.}
\newblock \bibinfo{journal}{\bibinfo{title}{{COMPASS}: {An} ab {initio}
  {force}-{field} {optimized} for {condensed}-{phase} {applications overview}
  with {details} on {alkane} and {benzene} {compounds}}}.
\newblock {\emph{\JournalTitle{The Journal of Physical Chemistry B}}}
  \textbf{\bibinfo{volume}{102}}, \bibinfo{pages}{7338--7364},
  \doiprefix\url{10.1021/jp980939v} (\bibinfo{year}{1998}).

\bibitem{shan_influence_2012}
\bibinfo{author}{Shan, M.} \emph{et~al.}
\newblock \bibinfo{journal}{\bibinfo{title}{Influence of chemical
  functionalization on the \ce{CO2}/\ce{N2} separation performance of porous
  graphene membranes}}.
\newblock {\emph{\JournalTitle{Nanoscale}}} \textbf{\bibinfo{volume}{4}},
  \bibinfo{pages}{5477}, \doiprefix\url{10.1039/c2nr31402a}
  (\bibinfo{year}{2012}).

\bibitem{wu_fluorine-modified_2014}
\bibinfo{author}{Wu, T.} \emph{et~al.}
\newblock \bibinfo{journal}{\bibinfo{title}{Fluorine-{modified} {porous}
  {graphene} as {membrane} for \ce{CO2}/\ce{N2} {separation}: {Molecular}
  {dynamic} and {first}-{principles} {simulations}}}.
\newblock {\emph{\JournalTitle{The Journal of Physical Chemistry C}}}
  \textbf{\bibinfo{volume}{118}}, \bibinfo{pages}{7369--7376},
  \doiprefix\url{10.1021/jp4096776} (\bibinfo{year}{2014}).

\bibitem{xu_insights_2015}
\bibinfo{author}{Xu, J.} \emph{et~al.}
\newblock \bibinfo{journal}{\bibinfo{title}{Insights into the \ce{H2}/\ce{CH4}
  {separation} {through} {two}-{dimensional} {graphene} {channels}: {Influence}
  of {sdge} {functionalization}}}.
\newblock {\emph{\JournalTitle{Nanoscale Research Letters}}}
  \textbf{\bibinfo{volume}{10}}, \bibinfo{pages}{492},
  \doiprefix\url{10.1186/s11671-015-1199-2} (\bibinfo{year}{2015}).

\bibitem{bu_first-principles_2013}
\bibinfo{author}{Bu, H.}, \bibinfo{author}{Zhao, M.}, \bibinfo{author}{Wang,
  A.} \& \bibinfo{author}{Wang, X.}
\newblock \bibinfo{journal}{\bibinfo{title}{First-principles prediction of the
  transition from graphdiyne to a superlattice of carbon nanotubes and graphene
  nanoribbons}}.
\newblock {\emph{\JournalTitle{Carbon}}} \textbf{\bibinfo{volume}{65}},
  \bibinfo{pages}{341--348}, \doiprefix\url{10.1016/j.carbon.2013.08.035}
  (\bibinfo{year}{2013}).

\bibitem{puigdollers_first-principles_2016}
\bibinfo{author}{Puigdollers, A.~R.}, \bibinfo{author}{Alonso, G.} \&
  \bibinfo{author}{Gamallo, P.}
\newblock \bibinfo{journal}{\bibinfo{title}{First-principles study of
  structural, elastic and electronic properties of $\alpha$-, $\beta$- and
  $\gamma$-graphyne}}.
\newblock {\emph{\JournalTitle{Carbon}}} \textbf{\bibinfo{volume}{96}},
  \bibinfo{pages}{879--887}, \doiprefix\url{10.1016/j.carbon.2015.10.043}
  (\bibinfo{year}{2016}).

\bibitem{li_be_2014}
\bibinfo{author}{Li, Y.}, \bibinfo{author}{Liao, Y.} \& \bibinfo{author}{Chen,
  Z.}
\newblock \bibinfo{journal}{\bibinfo{title}{\ce{Be2C} {monolayer} with
  {quasi}-{planar} {hexacoordinate} {carbons}: {A} {global} {minimum}
  {structure}}}.
\newblock {\emph{\JournalTitle{Angewandte Chemie International Edition}}}
  \textbf{\bibinfo{volume}{53}}, \bibinfo{pages}{7248--7252},
  \doiprefix\url{10.1002/anie.201403833} (\bibinfo{year}{2014}).

\bibitem{guo_co_2015}
\bibinfo{author}{Guo, H.} \emph{et~al.}
\newblock \bibinfo{journal}{\bibinfo{title}{\ce{CO2} {capture} on h--bn {sheet}
  with {high} {selectivity} {controlled} by {external} {electric} {field}}}.
\newblock {\emph{\JournalTitle{The Journal of Physical Chemistry C}}}
  \textbf{\bibinfo{volume}{119}}, \bibinfo{pages}{6912--6917},
  \doiprefix\url{10.1021/acs.jpcc.5b00681} (\bibinfo{year}{2015}).

\bibitem{song_graphenylene_2013}
\bibinfo{author}{Song, Q.} \emph{et~al.}
\newblock \bibinfo{journal}{\bibinfo{title}{Graphenylene, a unique
  two-dimensional carbon network with nondelocalized cyclohexatriene units}}.
\newblock {\emph{\JournalTitle{J. Mater. Chem. C}}}
  \textbf{\bibinfo{volume}{1}}, \bibinfo{pages}{38--41},
  \doiprefix\url{10.1039/C2TC00006G} (\bibinfo{year}{2013}).

\bibitem{zhu_theoretical_2017}
\bibinfo{author}{Zhu, L.} \emph{et~al.}
\newblock \bibinfo{journal}{\bibinfo{title}{Theoretical study of \ce{H2}
  separation performance of two-dimensional graphitic carbon oxide membrane}}.
\newblock {\emph{\JournalTitle{International Journal of Hydrogen Energy}}}
  \textbf{\bibinfo{volume}{42}}, \bibinfo{pages}{13120--13126},
  \doiprefix\url{10.1016/j.ijhydene.2017.04.043} (\bibinfo{year}{2017}).

\bibitem{brehm_travis_2011}
\bibinfo{author}{Brehm, M.} \& \bibinfo{author}{Kirchner, B.}
\newblock \bibinfo{journal}{\bibinfo{title}{{TRAVIS} - {A} {free} {Analyzer}
  and {visualizer} for {Monte} {Carlo} and {molecular} {dynamics}
  {trajectories}}}.
\newblock {\emph{\JournalTitle{Journal of Chemical Information and Modeling}}}
  \textbf{\bibinfo{volume}{51}}, \bibinfo{pages}{2007--2023},
  \doiprefix\url{10.1021/ci200217w} (\bibinfo{year}{2011}).

\bibitem{liu_insights_2013}
\bibinfo{author}{Liu, H.}, \bibinfo{author}{Dai, S.} \& \bibinfo{author}{Jiang,
  D.-e.}
\newblock \bibinfo{journal}{\bibinfo{title}{Insights into {CO2}/{N2} separation
  through nanoporous graphene from molecular dynamics}}.
\newblock {\emph{\JournalTitle{Nanoscale}}} \textbf{\bibinfo{volume}{5}},
  \bibinfo{pages}{9984}, \doiprefix\url{10.1039/c3nr02852f}
  (\bibinfo{year}{2013}).

\bibitem{wesolowski_pillared_2011}
\bibinfo{author}{Wesołowski, R.~P.} \& \bibinfo{author}{Terzyk, A.~P.}
\newblock \bibinfo{journal}{\bibinfo{title}{Pillared graphene as a gas
  separation membrane}}.
\newblock {\emph{\JournalTitle{Physical Chemistry Chemical Physics}}}
  \textbf{\bibinfo{volume}{13}}, \bibinfo{pages}{17027},
  \doiprefix\url{10.1039/c1cp21590f} (\bibinfo{year}{2011}).

\bibitem{du_separation_2011}
\bibinfo{author}{Du, H.} \emph{et~al.}
\newblock \bibinfo{journal}{\bibinfo{title}{Separation of {hydrogen} and
  {nitrogen} {gases} with {porous} {graphene} {membrane}}}.
\newblock {\emph{\JournalTitle{The Journal of Physical Chemistry C}}}
  \textbf{\bibinfo{volume}{115}}, \bibinfo{pages}{23261--23266},
  \doiprefix\url{10.1021/jp206258u} (\bibinfo{year}{2011}).

\end{thebibliography}
\end{document}